\documentclass[%
 aps,
nofootinbib,
 amsmath,amssymb,
 reprint,%
floatfix]{revtex4-2}

\usepackage{graphicx}

\usepackage{dcolumn}
\usepackage{bm}
\usepackage[utf8]{inputenc}
\usepackage[T1]{fontenc}
\usepackage{mathptmx}
\usepackage{etoolbox}
\usepackage{hyperref}

\usepackage{xcolor}
\definecolor{MainColor}{HTML}{686CB2}

\hypersetup{
	colorlinks,
 linkcolor=MainColor,
 citecolor=MainColor,
 urlcolor=MainColor
}

\usepackage[version=3]{mhchem}

\usepackage{comment}

\usepackage{cleveref}

\crefrangeformat{equation}{Eqs.~(#3#1#4) --~(#5#2#6)}

\usepackage[colorinlistoftodos]{todonotes}
\usepackage[font=Large,labelformat=simple,caption=false]{subfig}    

\usepackage{csquotes}

\makeatletter
\newlength{\sfp@hseplen}\newlength{\sfp@vseplen}
\define@cmdkey{subfigpos}[sfp@]{pos}[ul]{}
\define@cmdkey{subfigpos}[sfp@]{font}[\small]{}
\define@cmdkey{subfigpos}[sfp@]{vsep}[2\baselineskip]{\setlength{\sfp@vseplen}{\sfp@vsep}}
\define@cmdkey{subfigpos}[sfp@]{hsep}[10pt]{\setlength{\sfp@hseplen}{\sfp@hsep}}
\newcommand{\subfigimg}[3][,]{%
  \setkeys{Gin,subfigpos}{pos,font,vsep,hsep,#1}
  \setbox1=\hbox{\includegraphics{#3}}
  \ifnum\pdfstrcmp{\sfp@pos}{ul}=0
    \leavevmode\rlap{\usebox1}
    \rlap{\hspace*{\sfp@hsep}\raisebox{\dimexpr\ht1-\sfp@vsep}{\sfp@font{#2}}}
    \phantom{\usebox1}
  \else\ifnum\pdfstrcmp{\sfp@pos}{ur}=0
    \leavevmode\usebox1
    \llap{\raisebox{\dimexpr\ht1-\sfp@vsep}{\sfp@font{#2}}\hspace*{\sfp@hsep}}
  \else\ifnum\pdfstrcmp{\sfp@pos}{lr}=0
    \leavevmode\usebox1
    \llap{\raisebox{\sfp@vsep}{\sfp@font{#2}}\hspace*{\sfp@hsep}}
  \else
    \leavevmode\rlap{\usebox1}
    \rlap{\hspace*{\sfp@hseplen}\raisebox{\sfp@vsep}{\sfp@font{#2}}}
    \phantom{\usebox1}
  \fi\fi\fi
}

\makeatletter
\def\@email#1#2{%
 \endgroup
 \patchcmd{\titleblock@produce}
  {\frontmatter@RRAPformat}
  {\frontmatter@RRAPformat{\produce@RRAP{*#1\href{mailto:#2}{#2}}}\frontmatter@RRAPformat}
  {}{}
}%
\makeatother
\begin{document}

\preprint{AIP/123-QED}

\title[]{Asymmetric rectified electric fields for symmetric electrolytes}

\author{A. Barnaveli}
\author{R. van Roij}%
 \email{a.barnaveli@uu.nl}
\affiliation{ 
Institute for Theoretical Physics, Center for Extreme Matter and Emergent Phenomena, Utrecht University, Princetonplein 5, 3584 CC Utrecht, The Netherlands
}%

\date{\today}

\begin{abstract}

In this paper, building upon the discovery of asymmetric rectified electric fields (AREF) in recent experiments [S.H. Hashemi et al., Physical Review Letters {\bf 121}, 185504 (2018)], we explore the generation of AREF by applying a sawtooth-like voltage to 1:1 electrolytes with equal diffusion coefficients confined between two planar blocking electrodes. This differs from an earlier approach based on a sinusoidal AC voltage applied to 1:1 electrolytes with unequal diffusion coefficients. By numerically solving the full Poisson-Nernst-Planck equations, we demonstrate that AREF can be generated by a slow rise and a fast drop of the potential (or vice versa), even for electrolytes with equal diffusion coefficients of the cations and anions. We employ an analytically constructed equivalent electric circuit to explain the underlying physical mechanism. Importantly, we find that the strength of AREF can be effectively tuned from zero to its maximal value by only manipulating the time-dependence of the driving voltage, eliminating the necessity to modify the electrolyte composition between experiments. This provides valuable insights to control the manipulation of AREF, which facilitates enhanced applications in diverse electrochemical systems.

\end{abstract}

\maketitle

\section{\label{Sec:Intro} Introduction}

Studying the behavior of an aqueous electrolyte subjected to an externally applied oscillating electric field often involves the use of alternating current (AC) voltages. For instance, an AC voltage is commonly used in areas such as induced charge electrokinetics~\cite{bazant2004induced,bazant2009towards,ramos2016ac,squires2009induced,ramos1999ac}, particle assembly in electrolytes~\cite{hermanson2001dielectrophoretic,gierhart2007frequency,gangwal2008dielectrophoretic,gangwal2010programmed,smith2000electric,fan2005controllable,edwards2007electric,garcia2015self}, AC electroosmosis~\cite{bazant2010induced,song2017ac,du2019multifrequency,lee2018monte,talapatra2008double,chiou2008light,zhou2005lateral}, cyclic voltammetry~\cite{wang2013simulations,wang2012physical,zhao2023thermal,aderyani2021simulation,levey2023simulation,lin2022microscopic}, batteries~\cite{venkateswarlu1998ac,krichen2015ac,sheha2009ionic,pandey2014optimized,eswaragomathy2023preparation,li2008lithium,buvaneshwari2022preparation,tron2017surface}, sensing~\cite{song2017ac,lee2018monte,zhang2023iontronic}, and impedance spectroscopy~\cite{lasia2002electrochemical,barsoukov2018impedance,krichen2015ac,sheha2009ionic,pandey2014optimized,eswaragomathy2023preparation,li2008lithium,buvaneshwari2022preparation,tron2017surface}. One of the main reasons for choosing AC electric fields over DC fields in various applications is to avoid any net current or net charge in the system, since the field has a zero mean over one cycle. 

A basic geometry that can capture many of the essential physical effects of an AC field is a globally neutral 1:1 electrolyte of point-like ions confined between two blocking electrodes and subjected to a harmonic AC voltage. If the frequency of the AC voltage is relatively low or zero (as in equilibrium), then a so-called Electric Double Layer (EDL), consisting of the surface charges of the solid and a diffuse ionic cloud with opposite charge, will form at the interface between a charged solid (electrode, colloid, etc.) and an electrolyte. The EDL harbors a surplus of counter-ions and a reduced concentration of co-ions compared to the bulk, thereby screening the electric field of the electrode. The typical thickness of a fully formed EDL is equal to the Debye length $\lambda_D$, which is about $10$ nm for water with $1$ mM salt concentration at room temperature. One of the interesting recent findings in such a (vertical) system with horizontal electrodes concerned colloids floating in the gravitational field. Here, charged colloids suspended in an aqueous electrolyte were confined between two horizontal blocking electrodes that were driven by a harmonic AC potential. Contrary to intuition, rather than sedimenting in the gravitational field the colloidal particles were observed to float against the gravitational pull \cite{woehl2015bifurcation, bukosky2015simultaneous}. This led to a theoretical investigation to elucidate the source of the force that allows the colloids to withstand the gravitational field. In Ref.~\cite{hashemi2018oscillating} it was shown that period-averaged electrode charge is not necessarily zero in the case of cations (+) and anions (-) with unequal diffusion coefficients,  $D_+\neq D_-$. The resulting period-averaged  induced electric field is therefore also non-zero and stretches from the electrodes well into the bulk of the electrolyte. It was termed Asymmetric Rectified Electric Field (AREF). The electric force generated by AREF was proposed as a mechanism that would enable the colloids to counteract the gravitational pull. It is noteworthy that a recent study has proposed alternative mechanisms for colloidal floating, including dielectrophoresis (DEP) or electrohydrodynamical (EHD) mechanisms \cite{chen2023dielectrophoretic}. Interestingly, the predominant contribution of each mechanism to the floating height of colloids remains a subject of investigation.

\begin{figure} [ht]
\captionsetup[subfigure]{labelformat=empty}
\subfloat[\label{Fig:SawTooth}]{\subfigimg[width=\linewidth,pos=tl,vsep=4.4cm,hsep=1.4cm]{\large (a)}{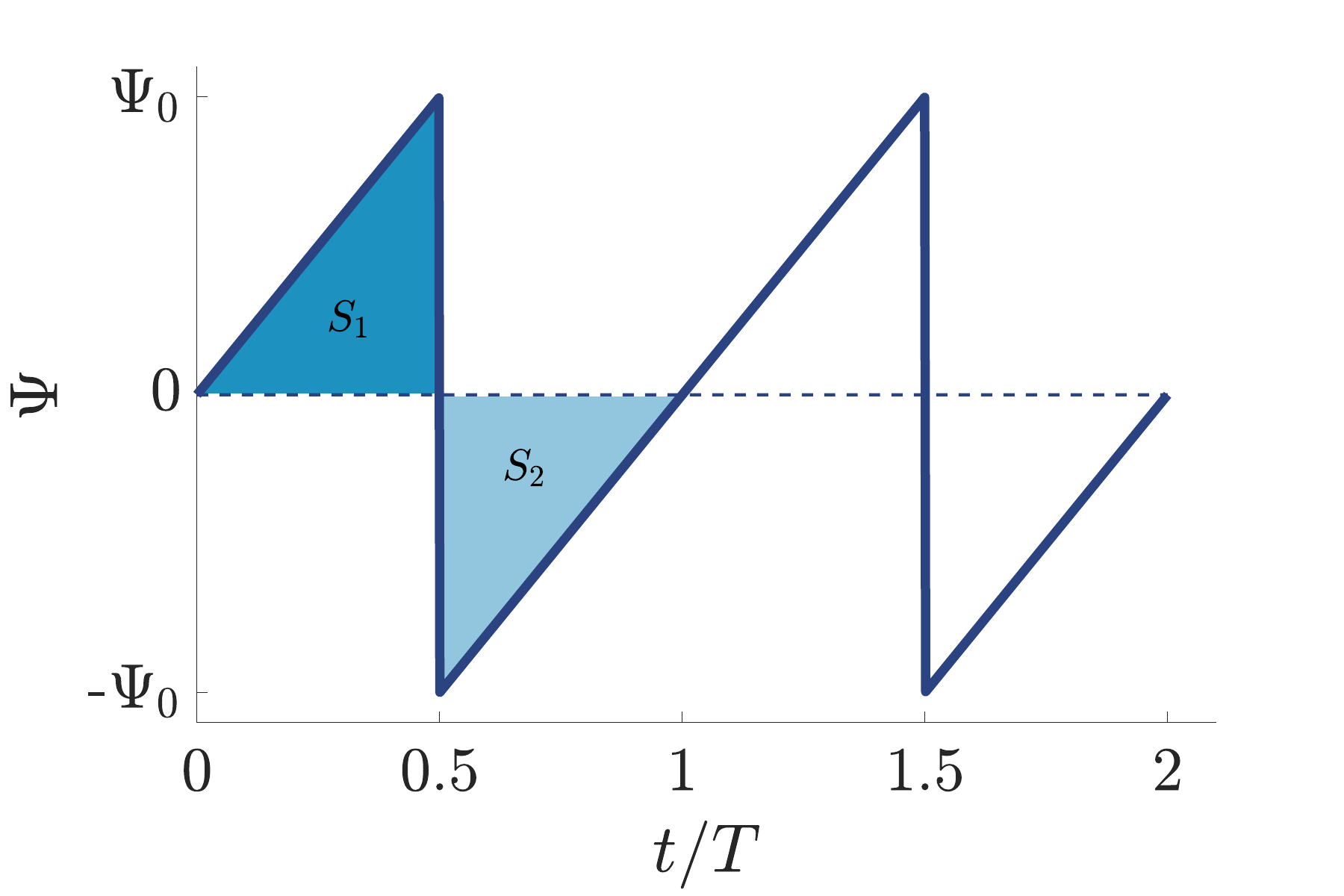}}\quad
\subfloat[\label{Fig:SawTooth2Term}]{\subfigimg[width=\linewidth,pos=tl,vsep=4.4cm,hsep=1.4cm]{\large (b)}{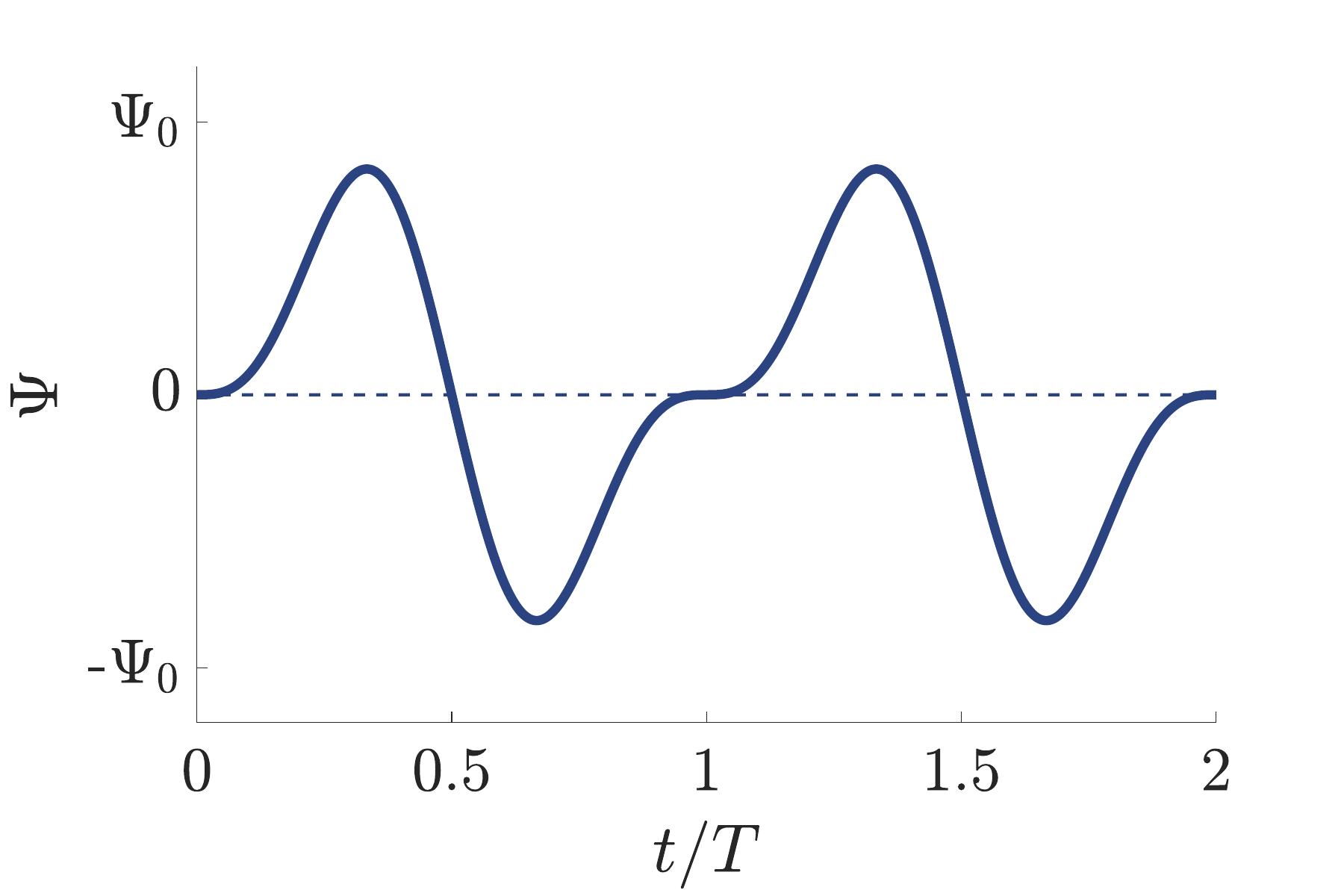}}
\caption{Two periods of (a) the full sawtooth voltage $\Psi(t)$ based on Eq.~\eqref{eq:SawTooth} and (b) the two-term sawtooth voltage based on Eq.~\eqref{eq:Sawtooth2Term}. Both voltages have a period $T$, feature an asymmetry between (slow) rising and (fast) lowering voltages, and average out to zero during a period. The two-term sawtooth avoids sharp transitions, rendering itself more convenient for numerical calculations.}
\end{figure}

This paper focuses on AREF. The authors of the original study extensively explored AREF from sinusoidal voltages by examining its space dependence on various system parameters in Ref.~\cite{hashemi2019asymmetric}, numerically solving the governing system of non-linear differential equations in Ref.~\cite{hashemi2020perturbation}, and investigating the application of AREF in reversing the flow of electroosmosis in Ref.~\cite{hashemi2020asymmetric}. Nevertheless, several aspects of the underlying physical mechanism of AREF remained unclear. In our recent publication \cite{barnaveli2024asymmetric} we employ equivalent electric circuits to devise a simplified toy model that qualitatively reproduces the parameter dependencies of AREF, shedding light on the underlying physical mechanism. It was explained how the asymmetry of ion diffusion coefficients in the electrolyte can create AREF. However, the scope of manipulating AREF is constrained by the rather limited range of ion diffusion coefficients and their disparities. Furthermore, experimental studies on AREFs necessitate altering electrolytes for each new experiment, demanding a significant investment of time and effort. To address these challenges, we opted to study one and the same electrolyte, for simplicity a symmetric 1:1 electrolyte with equal ion diffusion coefficients $D\equiv D_+=D_-$,  and instead study the possibility of introducing the necessary asymmetry for AREF generation through the functional form of the driving potential. A convenient form that is both asymmetric and periodic, yet averages to zero over time, is the so-called ``sawtooth'' potential 
\begin{equation} \label{eq:SawTooth}
	{
		\Psi(t)=\frac{2\Psi_0 }{\pi}\sum_{n=1}^\infty (-1)^{n+1}\frac{\sin{(n\omega t)}}{n},
	}
\end{equation}
where $\Psi_0>0$ is the amplitude and $T=2\pi/\omega$ the period the of the driving voltage $\Psi(t)$. In Fig.~\ref{Fig:SawTooth}, where we plot two periods of $\Psi(t)$ given by Eq.~\eqref{eq:SawTooth} as a function of the dimensionless time $t/T$ , we see that the sawtooth function rises steadily towards its maximum $\Psi_0$ and then drops ``instantaneously'' to its  minimum $-\Psi_0$. This slow rise and fast drop breaks the symmetry of the charging and discharging processes at the electrodes, as we will see. At the same time, the (absolute) areas $S_1$ and $S_2$ under the curve are equal, $S_1=S_2$, resulting in a period-averaged applied potential equal to zero, i.e. there is no direct bias of the voltage.  

While the full sawtooth function is indeed a very convenient candidate for the time-dependence of the driving voltage, it is less attractive for the numerical study that we undertake in this work, not only because of the large number of required harmonic ``modes'' in Eq.~\eqref{eq:SawTooth} but also because of the discontinuity of the full potential.  It turns out that the essence of the creation mechanism of AREF can be studied in full detail by avoiding the sharpest feature of the full potential and keeping only the first two terms in the sawtooth series of Eq.~\eqref{eq:SawTooth}. Thus, henceforth the driving voltage of interest is given by 
\begin{equation} \label{eq:Sawtooth2Term}
	{
		\Psi(t)=\frac{2\Psi_0 }{\pi}\sum_{n=1}^2 (-1)^{n+1}\frac{\sin{(n\omega t)}}{n},
	}
\end{equation}
which is plotted in Fig.~\ref{Fig:SawTooth2Term}. One checks that the role of the second harmonic term is to break the symmetry between rising and lowering voltages. All numerical results in this paper will be based on this ``two-term'' sawtooth function, that captures the key physics even though its actual amplitude is only $\sim0.9\Psi_0$.   However, for convenience and clarity we will refer to the full sawtooth function when explaining and discussing the AREF mechanism.

This paper is structured as follows: In Section~\ref{Sec:PNP}, we present the system of interest along with the Poisson-Nernst-Planck (PNP) equations that control the processes in the electrolytic cell. In Section~\ref{Sec:AREF} we explain how AREF effects are generated under the influence of the sawtooth driving potential for a specific set of system parameters. In Section~\ref{Sec:ParametricStudy}, we use numerical methods to investigate how the AREF varies with the main system parameters. Finally, in Section~\ref{Sec:Conclusion} we sum up and discuss our results.


\section{Poisson-Nernst-Planck Equations}
\label{Sec:PNP}

\begin{figure} [ht]
	\centering
	\includegraphics[width=0.99\linewidth]{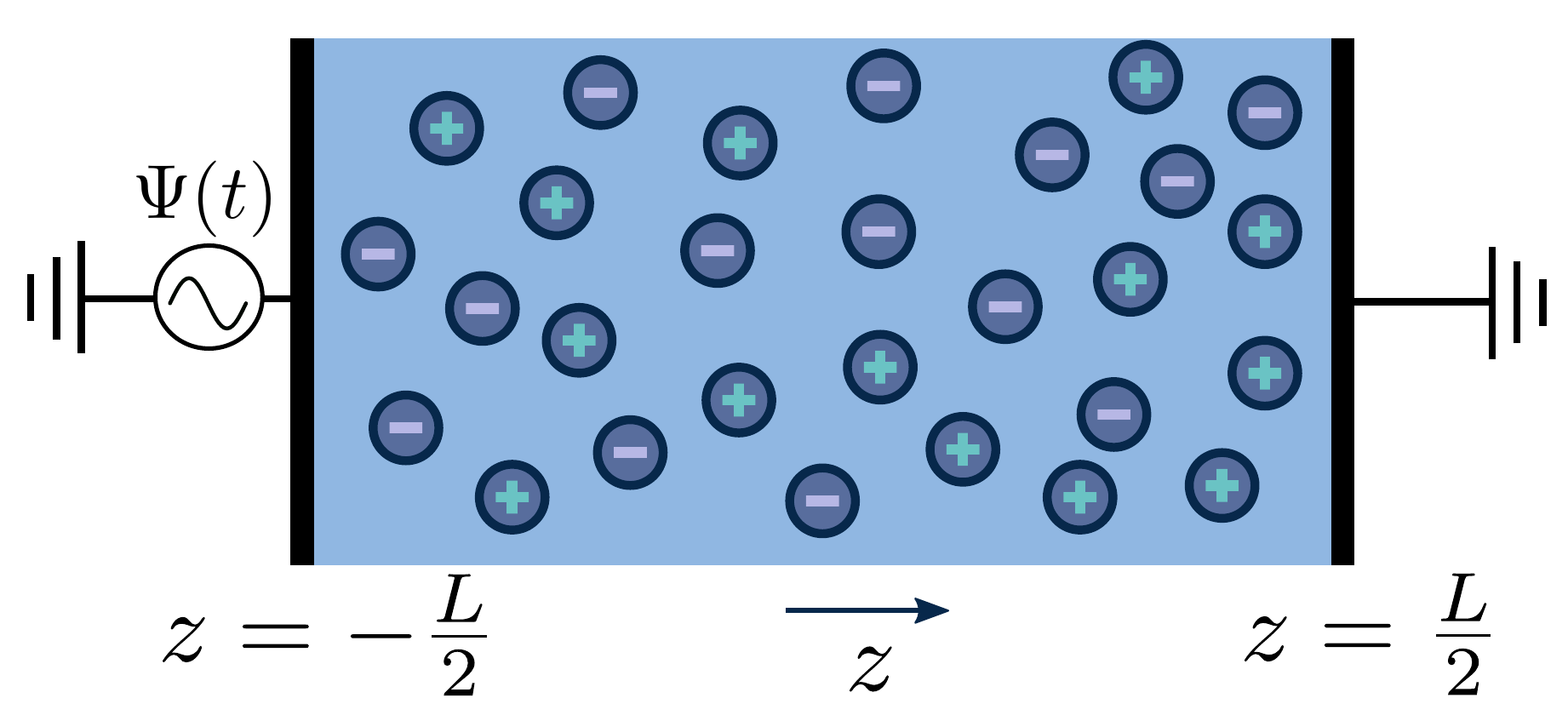}
	\caption{Schemtaic illustration of the aqueous 1:1 electrolyte under consideration, comprising a continuous solvent and two ionic species, enclosed between two parallel blocking electrodes with a separation distance $L$. The ions in the electrolyte are driven by the time-dependent electric sawtooth potential $\Psi(t)$ of Eq.~\eqref{eq:Sawtooth2Term} applied to the electrode at $z=-\frac{L}{2}$, while the opposite electrode at $z=\frac{L}{2}$ remains grounded.} \label{Fig:TwoPlatesSetupAREF}
\end{figure}

The system of interest, schematically illustrated in Fig.~\ref{Fig:TwoPlatesSetupAREF}, is essentially the same electrolytic cell as the one considered in our previous paper \cite{barnaveli2024asymmetric}, therefore its description and the notation we use will follow Ref.~\cite{barnaveli2024asymmetric} very closely. The cell comprises a three-dimensional aqueous electrolyte with a relative dielectric constant $\epsilon$ at room temperature, confined between two parallel macroscopic planar electrodes separated by a distance $L$. We assume translational symmetry in the lateral directions. Apart from the continuum solvent, the electrolyte is composed of two types of monovalent point-like ions: cations $(+)$ and anions $(-)$ with valencies $\pm 1$ and equal diffusion coefficients $D_\pm \equiv D$. The total number of cations and anions is equal, ensuring overall electroneutrality in the system. The electrodes are blocking, preventing ions from leaving the electrolyte, and we exclude any chemical REDOX reactions. The system is subjected to the AC sawtooth voltage of Eq.~\eqref{eq:Sawtooth2Term} containing only two terms in the series, applied to the left electrode placed in the plane $z=-\frac{L}{2}$, whereas the right one, situated at $z=\frac{L}{2}$, remains grounded. The imposed angular frequency is denoted by $\omega$, and $\Psi_0$ represents the amplitude of the applied voltage. 

We study this system in terms of the the Poisson-Nernst-Planck (PNP) equations. The ionic fluxes, denoted as $J_\pm(z,t)$, comprise a diffusive component arising from ion concentration gradients and a conductive component resulting from the potential gradient. These aspects are collectively described by the Nernst-Planck equation given by
\begin{equation} \label{eq:NernstPlanckAREF}
	{		J_{\pm}(z,t)=-D\bigg(\frac{\partial c_\pm(z,t)}{\partial z} \pm \beta e c_\pm(z,t)\frac{\partial \Psi(z,t)}{\partial z}\bigg),
	}
\end{equation}
where $c_\pm(z,t)$ represent the concentrations of cations ($+$) and anions ($-$) at the position $z$ and time $t$ and $\Psi(z,t)$ the local electrostatic potential. Here $e$ is the elementary charge and $\beta^{-1}$ the product of the Boltzmann constant and the temperature. Eq.~\eqref{eq:NernstPlanckAREF} also assumes spatially constant diffusion coefficients. Given the absence of chemical reactions in the system, the concentrations and fluxes are connected through the continuity equation
\begin{equation} \label{eq:ContinuityAREF}
	{		\frac{\partial c_\pm (z,t)}{\partial t}+\frac{\partial J_\pm(z,t)}{\partial z}=0.
	}
\end{equation}
The local potential profile $\Psi(z,t)$ is connected to the local charge density $e \big(c_+(z,t)- c_-(z,t)\big)$ through the Poisson equation, which for $|z|<\frac{L}{2}$ reads
\begin{equation} \label{eq:PoissonAREF}
	{\frac{\partial ^2 \Psi(z,t)}{\partial z^2}=-\frac{e}{\epsilon_0\epsilon}\big(c_+(z,t)- c_-(z,t)\big),
	}
\end{equation}
where $\epsilon_0$ is the permittivity of vacuum and $\epsilon=80$ represents water as a structureless continuum. 

The PNP equations \eqref{eq:NernstPlanckAREF}, \eqref{eq:ContinuityAREF}, and \eqref{eq:PoissonAREF} form a closed set that fully describes the time-dependent profiles of the concentrations $c_\pm$, the fluxes $J_\pm$, and the potential $\Psi$. The explicit solution of the PNP equations requires boundary and initial conditions, for which we take
\begin{align} 
	\label{eq:AREF_Boundary_Conditions_Poisson_1}
	\Psi(-L/2,t)&=\frac{2\Psi_0 }{\pi}\bigg(\sin{(\omega t)}-\frac{1}{2}\sin{(2\omega t)}\bigg),\vspace{0.5cm
	}\\ 
    \label{eq:AREF_Boundary_Conditions_Poisson_2}
	\Psi(L/2,t)&=0,\\
	\label{eq:AREF_Boundary_Conditions_NP}
	J_\pm(-L/2,t)&=J_\pm(L/2,t)=0, \\
	\label{eq:AREF_Boundary_Conditions_cbulk}
	c_\pm(z,t=0)&=c_s \hspace{0.2cm} \text{for } \hspace{0.2cm} z\in[-L/2,L/2].
\end{align}
Here $c_{s}$ represents the constant initial salt concentration, which is identical for both ionic species in the $1:1$ electrolyte of interest here and thus satisfies global charge neutrality. As implied by Eq.~\eqref{eq:ContinuityAREF} coupled with the boundary conditions specified in Eq.~\eqref{eq:AREF_Boundary_Conditions_NP}, the total number of anions and cations in the system is conserved such that
\begin{equation} \label{eq:AREFContinuity}
	{
		\frac{1}{L}\int_{-L/2}^{L/2}c_\pm(z,t)\,d z=c_s 
	}
\end{equation}
is satisfied at all times $t\geq 0$. For a given set of parameters $\Psi_0$, $\omega$, $D$, $c_s$, and $L$ \Crefrange{eq:PoissonAREF}{eq:AREFContinuity} constitute the system of non-linear coupled differential equations. We employ the finite-element solver of COMSOL\textregistered \, to numerically solve these equations.

Convenient insight into relevant dimensionless system parameters can be obtained as follows. In the static (low-frequency) limit equilibrium holds, such that the applied potential $\Psi(-L/2,t)=\Psi_0$ is a time-independent constant and $J_\pm (z,t)=0$. In the linear-screening regime with $|\beta e \Psi_0|\lesssim 1$, the EDLs get fully developed at the two electrodes and the NP equation \eqref{eq:NernstPlanckAREF} can be integrated to obtain the Boltzmann distribution
\begin{equation} \label{eq:Boltzmann_Distribution}
    {
    c_\pm(z)=c_s'\bigg( 1 \mp  \frac{\Psi_0 \beta e \sinh{(\kappa z)}}{2\sinh{(\kappa L/2)}}\bigg),
    }
\end{equation}
with $\kappa^{-1}$ the characteristic Debye length of the equilibrium EDL given by
\begin{equation} \label{eq:Debye_Length}
    {
    \kappa^{-1}=\sqrt{\frac{\epsilon \epsilon_0}{2 e^2 \beta c'_s}}\equiv \lambda_D.
    }
\end{equation}
The concentration $c_s'$ is an integration constant that is very close to $c_s$ in the large $L$-limit of interest here, so throughout the paper we set $c'_s=c_s$ in the definition of $\lambda_D$. In this limit, as we have shown before in Ref.~\cite{barnaveli2024asymmetric}, the characteristic timescale of EDL formation \cite{bazant2004diffuse} is written as the $RC$ time
\begin{equation} \label{eq:tau_RC}
    {
    \tau_{RC} = \frac{L \lambda_D }{2 D}=\frac{L}{2\kappa D}.
    }
\end{equation}
For future convenience we also define the Debye time
\begin{equation} \label{eq:tau_D}
    {
    \tau_D = \frac{1}{\kappa^2 D}=\tau_{RC}\frac{2}{\kappa L},
    }
\end{equation}
during which the ions diffuse over a  distance of the order of the Debye length \cite{bazant2004diffuse,rubinstein2009reexamination}.

For the convenience of numerical investigation of AREF, we establish a standard parameter set that includes the (dimensionless) amplitude and frequency of the driving potential, denoted as $\beta e \Psi_0=3$ and $\omega \tau_{RC}=1$, respectively. The standard (dimensionless) system size is fixed at $\kappa L=50$. We note that this standard parameter set is physically realistic, as it corresponds for an aqueous 1:1 electrolyte with a salt concentration $c_s=1~\text{mM}$ to a Debye length $\lambda_D=10~\text{nm}$ and hence a system length $L=500~\text{nm}$, and with a typical diffusion coefficient $D=1.09~\text{$\mu$m$^2$/ms}$ we find $\tau_{RC}=2.3~\text{$\mu$s}$ and hence a driving period $T=14.4~\text{$\mu$s}$. Any deviation from this standard set will be explicitly stated. All measurements are performed in the late-time limit-cycle regime, when all the transient effects have vanished. This way all the time-dependencies in the system have the same period as that of the driving voltage, with at most a phase difference as we will see.

\section{AREF From SawTooth Potentials}
\label{Sec:AREF}

Most of the previous work on AREF concentrated on asymmetric electrolytes containing ions with unequal diffusion coefficients driven by a harmonic (single-frequency sinusoidal) voltage \cite{bukosky2015simultaneous,woehl2015bifurcation,hashemi2018oscillating,hashemi2019asymmetric,hashemi2020asymmetric,hashemi2020perturbation,chen2023dielectrophoretic,balu2022electrochemical}. To appreciate the differences of AREF between these asymmetric electrolytes and the systems of interest here consisting of a symmetric 1:1 electrolyte (with equal diffusion coefficients) driven by the sawtooth potential of Eq.~\eqref{eq:AREF_Boundary_Conditions_Poisson_1}, we briefly recall the mechanism of AREF in the asymmetric case.  

As was discussed in Ref.~\cite{barnaveli2024asymmetric}, the mechanism behind the creation of AREF in a system with an asymmetric electrolyte relies on the concentration \emph{difference} of the faster (more mobile) ions gathering at the electrodes during a half-period $T/2$ and the slower (less mobile) oppositely charged ions during the complementary half period, an effect that is particularly strong for intermediate driving frequencies $\omega \tau_{RC}\sim 1$. As a result, in the vicinity of both electrodes the period-averaged concentration of the faster ions exceeds that of the slower ions, and the resulting period-averaged charge distribution $e\langle c_+-c_-\rangle(z)$ in the electrolyte was found to be nonzero and results in a nontrivial period-averaged electrostatic potential $\langle\Psi\rangle(z)$ and an associated period-averaged electric field (AREF) $\langle E\rangle(z) =-d\langle \psi\rangle(z)/dz$. Here we defined the period-average of a function $f(z,t)$ as 
\begin{equation} \label{eq:Time_Average_Def}
	\langle f \rangle(z)=\frac{1}{T}\int_{t_0}^{t_{0}+T}  f(z,t) \,dt,
\end{equation}
where $t_{0}$ is the (sufficiently late) time at which we start averaging. Because of the symmetry and equivalence between the two electrodes, at least at the period-averaged level, we find (for the asymmetric electrolyte with sinusoidal driving) perfect mirror symmetry with respect to the midplane for the period-averaged potential, so $\langle\Psi\rangle(z)=\langle\Psi\rangle(-z)$, and likewise for the ionic concentrations and the charge density. The electric field, by contrast, exhibits perfect anti-mirror symmetry with respect to the midplane, thus $\langle E\rangle(z)=-\langle E\rangle(-z)$ \cite{barnaveli2024asymmetric}. As a consequence of this symmetry, it was found in Ref.~\cite{barnaveli2024asymmetric} that a convenient integral quantity to characterize (the strength of) AREF was the time- and space-averaged (dimensionless) electric potential $U\equiv \beta e \frac{1}{L}\int_{-\frac{L}{2}}^{\frac{L}{2}}\,dz \langle \Psi \rangle(z)$. An additional consequence of these (anti-)symmetries combined with global charge neutrality was a vanishing period-averaged surface charge density $\langle\sigma\rangle$ on both electrodes at $z=\pm L/2$, such that not only $\langle\Psi\rangle(\pm L/2)=0$ but also  $\langle E\rangle(\pm L/2)=0$ for asymmetric electrolytes with symmetric driving voltages.

\begin{figure} [ht!]
\captionsetup[subfigure]{labelformat=empty}
\subfloat[\label{Fig:AvgCpCn}]{\subfigimg[width=\linewidth,pos=tl,vsep=4.2cm,hsep=1.25cm]{\large (a)}{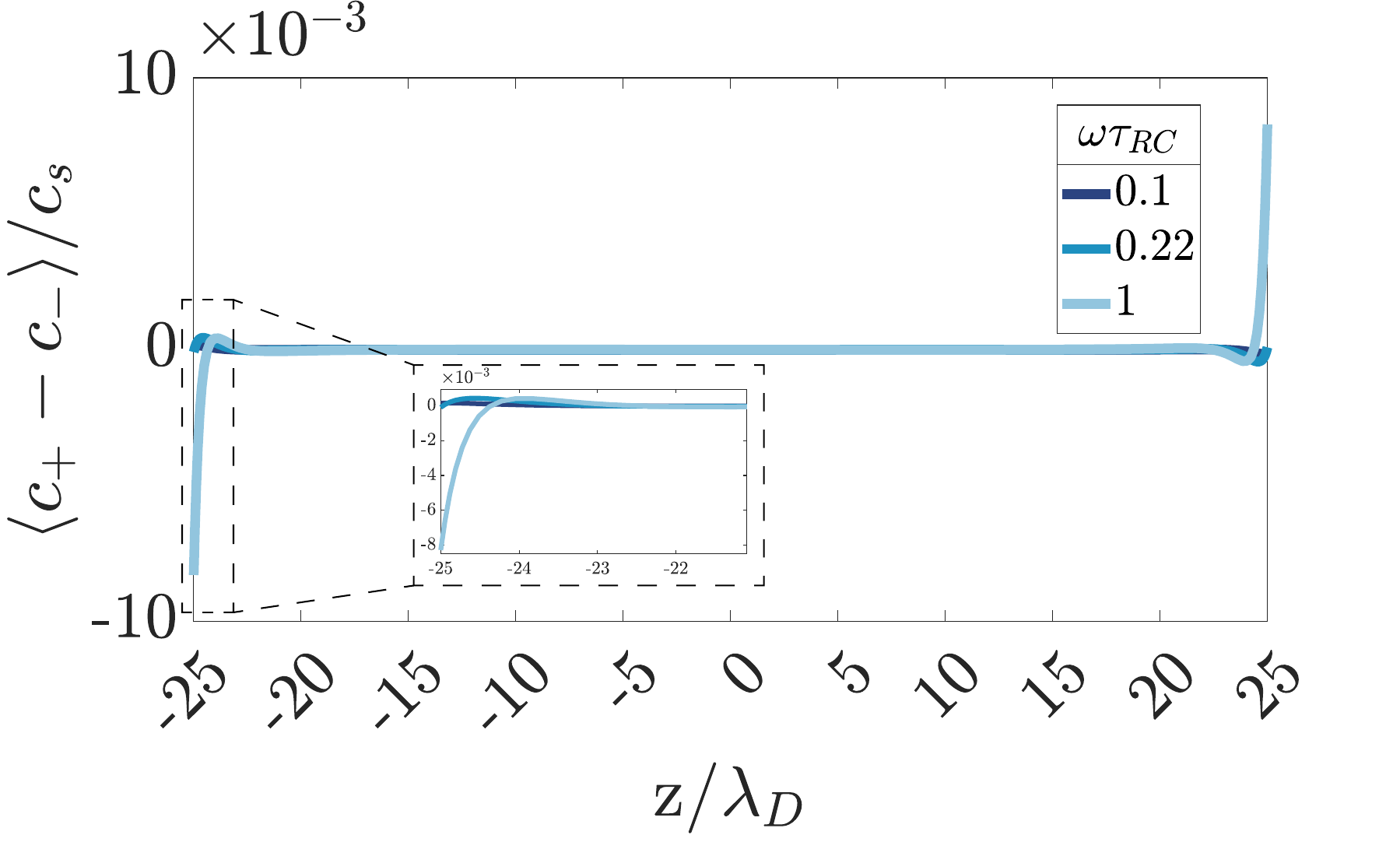}}\quad
\subfloat[\label{Fig:AvgE}]{\subfigimg[width=\linewidth,pos=tl,vsep=1.6cm,hsep=1.35cm]{\large (b)}{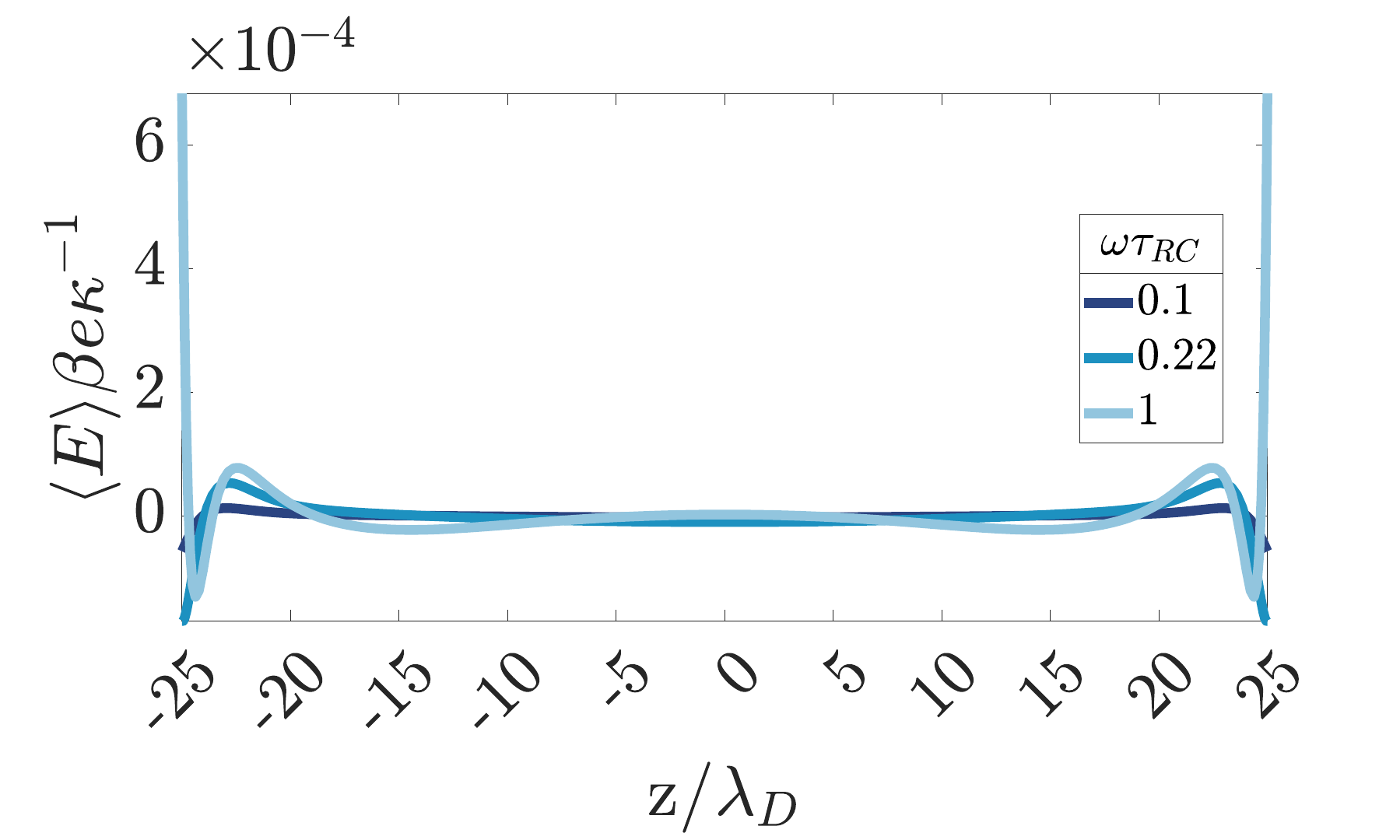}}\quad
\subfloat[\label{Fig:AvgVoltage}]{\subfigimg[width=\linewidth,pos=tl,vsep=1.5cm,hsep=1.15cm]{\large (c)}{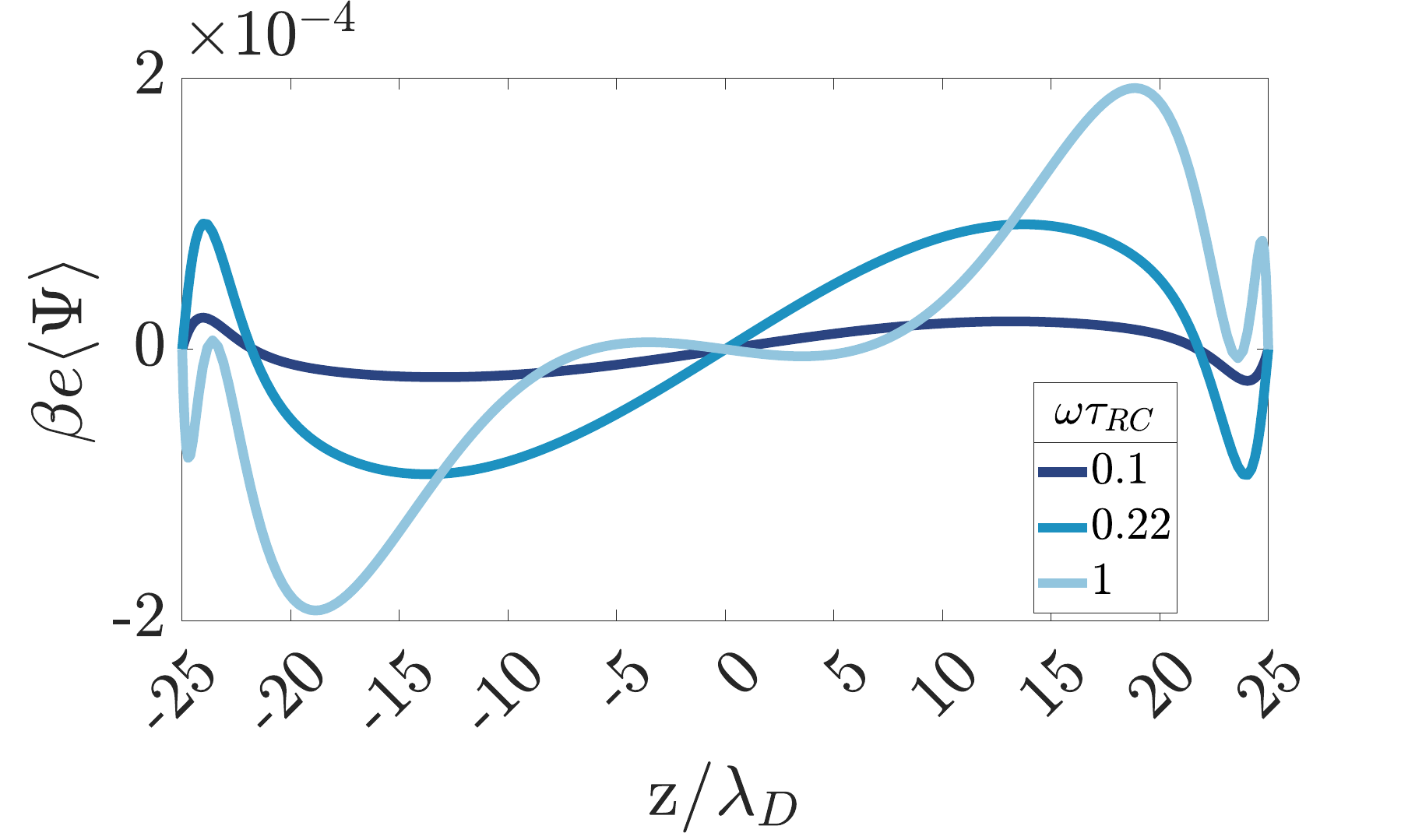}}
\caption{Time-averaged dimensionless spatial profiles of the (a) ionic charge density $\langle c_+-c_-\rangle/c_s$, (b) electric field $\beta e\kappa^{-1}\langle E\rangle$, and (c) electric potential $\beta e \langle \Psi \rangle$ in a  1:1 aqueous electrolyte confined between two planar electrodes separated by distance $L=50\lambda_D$. The electrode at $z=L/2$ is grounded, whereas the one at $z=-L/2$ is driven by an AC Sawtooth potential of Eq.~\eqref{eq:AREF_Boundary_Conditions_Poisson_1} with amplitude $\Psi_0=3/\beta e=75$ mV. Three different driving frequencies $\omega\tau_{RC}=0.1,0.22,1$ with $RC$-time $\tau_{RC}$ given by Eq.~\eqref{eq:tau_RC} are denoted with different colors. \label{Fig:AREF_AVERAGES}}
\end{figure}

Compared to the case of asymmetric ion diffusion coefficients that we just discussed, the system of a 1:1 electrolyte with equal ionic diffusion coefficients driven by the asymmetric sawtooth voltage has a different mechanism for AREF creation. This is immediately apparent from Fig.~\ref{Fig:AvgCpCn}, that shows the numerical solution of the PNP equations of the period-averaged charge density profile $\langle c_+-c_-\rangle(z)$ for our standard parameter set. At the left electrode placed at $z=-L/2$ we see a period-averaged accumulation of negative ionic charge, whereas on the opposite side at $z=L/2$ an equal but opposite (positive) charge density accumulates in the vicinity of the electrode. Clearly, this charge density profile is anti-symmetric with respect to mirroring in the midplane, $\langle c_+-c_-\rangle(z)=-\langle c_+-c_-\rangle(-z)$, which contrasts the mirror symmetry we encountered earlier in the cases of unequal ionic mobilities. Such an antisymmetric period-averaged charge distribution creates a perfectly mirror-symmetric AREF $\langle E(z)\rangle$, as also shown in Fig.~\ref{Fig:AvgE}, where we notice that the electric fields at $z=\pm L/2$, so at the electrodes, do not vanish.  This implies by the Gauss law that the period-averaged surface charge $\langle \sigma \rangle$ on the electrodes is non-zero in this case. At the same time we see in Fig.~\ref{Fig:AvgVoltage} that the period-averaged potential profile $\langle \Psi \rangle(z)$ follows the anti-mirror-symmetry of the charge distribution.  As a consequence, its spatial average $U$ will be identically zero, which implies that, unlike in Ref.~\cite{barnaveli2024asymmetric}, it cannot be used as a measure for the AREF strength. Instead, it is now natural to use the time-averaged surface charge density $\langle \sigma \rangle$ on the electrodes for this purpose, or rather its dimensionless version 
\begin{equation} \label{eq:Sigma_Defined}
 \sigma' \equiv \frac{\langle \sigma \rangle}{\sigma_m} = \frac{\beta e \kappa^{-1} \langle E\rangle(z)}{4\pi \sinh(\beta e \Psi_0/2)}\Big|_{z=-L/2} ,
\end{equation}
where we introduced the Gouy-Chapman surface charge density 
$\sigma_m=e(\kappa/\lambda_B)\sinh(\beta e\Psi_0/2)\approx 7.6~\text{mC/m}^2$ at the static voltage $\beta e \Psi_0=3$ as a reference, with the Bjerrum and Debye length set to $\lambda_B=e^2/4\pi\epsilon_0\epsilon k_BT\simeq 0.72~\text{nm}$ and $\kappa^{-1}\simeq 10~\text{nm}$, respectively. 

\begin{figure}
\captionsetup[subfigure]{labelformat=empty}
\subfloat[\label{Fig:Equivalent_Circuit_Full}]{\subfigimg[width=0.51\linewidth,pos=tl,vsep=3.2cm,hsep=1.2cm]{\large (a)}{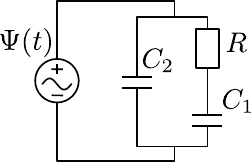}}\quad
\subfloat[\label{Fig:Equivalent_Circuit_Simplified}]{\subfigimg[width=0.45\linewidth,pos=tl,vsep=3.2cm,hsep=1.2cm]{\large (b)}{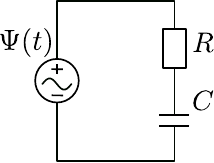}}
\caption{(a) equivalent electric circuit corresponding to the large electrolytic cell with $L\gg \kappa^{-1}$ in the linear regime. Resistance and capacitance of the cell at infinite frequency are denoted by $R$ and $C_2$ respectively, whereas the total capacitance of two fully developed electric double layers at the electrodes is denoted by $C_1$, as described by Eqs.~\eqref{eq:C_DL_Infty} and \eqref{eq:R_Infty}. (b) Simplified equivalent electric circuit corresponding to the low-frequency case $\omega \tau_{RC}\ll \sqrt{\kappa L/2}$ with $C=C_1$.}
\end{figure}

To understand the mechanism behind AREF in the present system, we will use the so-called equivalent circuit corresponding to the system that we are studying. It is well known that several aspects of electrolytic systems can often be approximated by equivalent electronic circuits \cite{macdonald1953theory,macdonald1971electrical,barker1966equivalent,abouzari2009physical,cole1941dispersion}, with Ref.~\cite{geddes1997historical} providing a historical overview on this matter. As was shown in Ref.~\cite{barnaveli2024asymmetric},  the system in Fig.~\ref{Fig:TwoPlatesSetupAREF} can in the linear screening regime $\beta e \Psi_0 \ll 1$ be approximated by the circuit shown in Fig.~\ref{Fig:Equivalent_Circuit_Full},
where the capacitors $C_1$ and $C_2$ and the resistor $R$ take, for an electrolytic system of lateral area $A$, the form 
\begin{align} 
    \label{eq:C_DL_Infty}
    C_1&= \frac{A \epsilon \epsilon_0 \kappa }{2}; \hspace{0.8cm}  C_2= \frac{A \epsilon \epsilon_0 }{L}, \\
    \label{eq:R_Infty}
    R&= \frac{L}{A}\cdot \frac{1}{ \epsilon \epsilon_0 D \kappa^2 }=\frac{L}{A}\cdot \frac{1}{2 D \beta e^2 c_s }.
\end{align}
Physically $R$ corresponds to the Ohmic resistance of the homogeneous aqueous electrolyte with monovalent charge carriers of concentration $2c_s$ and mobility $\beta D$, and $C_1$ represents the capacity of the EDLs at the electrodes - it is the net capacity of the two fully developed EDLs in series, each with the linear-screening capacitance  $A\epsilon \epsilon_0\kappa$. Similarly, $C_2$ represents the purely dielectric capacitance of a 
water-filled parallel-plate capacitor without any ionic charge carriers and characterized by the size $L$ and area $A$. 

\begin{figure} [ht]
	\centering
	\includegraphics[width=0.99\linewidth]{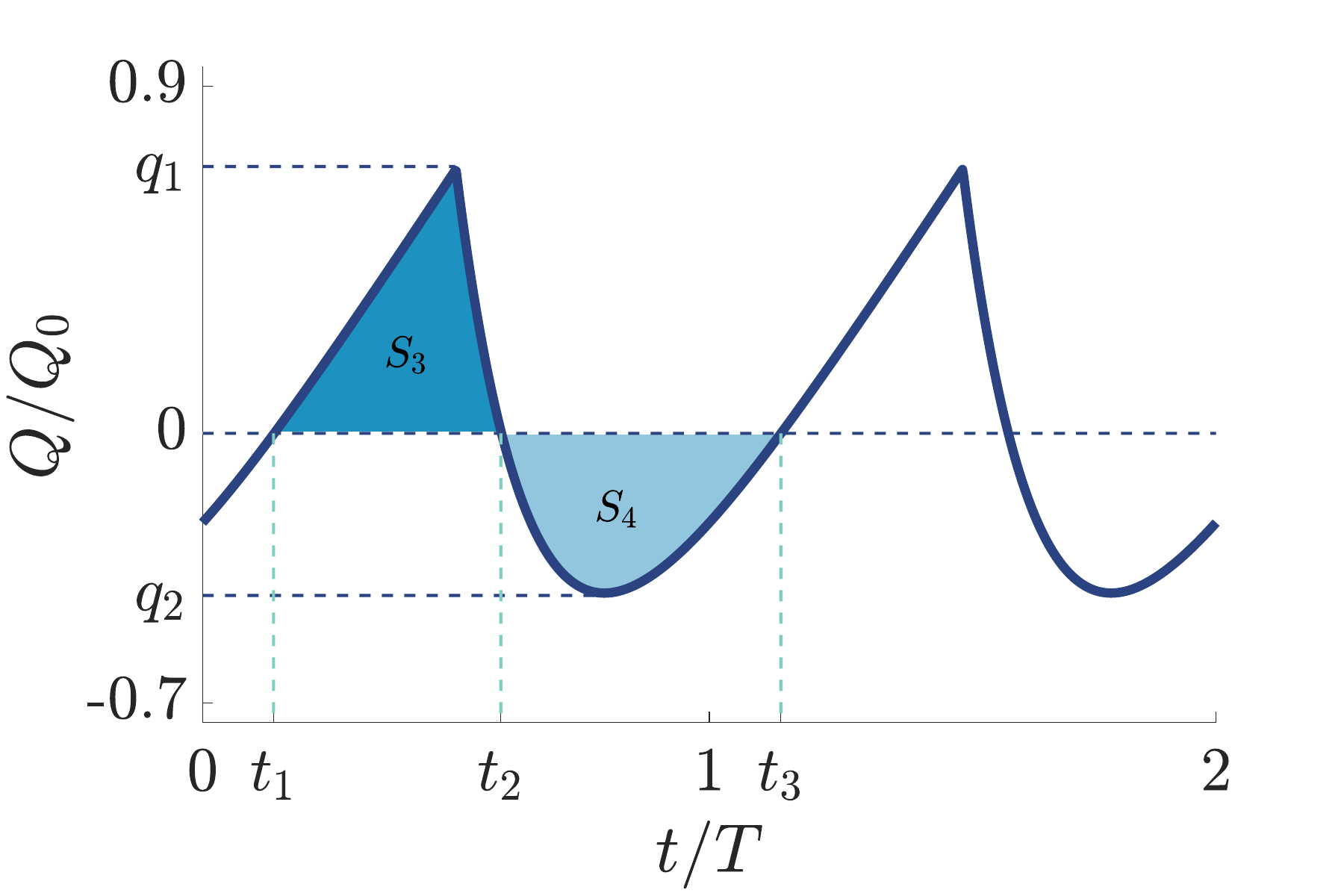}
	\caption{The time-dependent charge $Q(t)$ (in units of $Q_0$) as defined in Eq.~\eqref{eq:ChargeDynamics} stored in the capacitor of the linear equivalent circuit of Fig.~\ref{Fig:Equivalent_Circuit_Simplified} as a function of time for the full sawtooth potential $\Psi(t)$ of Eq.~\eqref{eq:SawTooth}. The asymmetry in the driving potential introduces not only an asymmetry of the positive and negative charge amplitudes, $|q_1|\neq |q_2|$, but also of the time interval that the charge is positive or negative, $t_2-t_1\neq t_3-t_2$. For linear circuits, or linear screening, this translates into a vanishing period-averaged charge since $S_3=S_4$ identically. In the non-linear case of the electrolytic cell at high voltages, however, this condition gets violated and results in a non-zero period-averaged surface charge $\sigma'$ on the electrodes.} \label{Fig:ChargeDynamics}
\end{figure}

Despite the circuit of Fig.~\ref{Fig:Equivalent_Circuit_Full} being only a quantitative mapping in the case of the linearized PNP equations valid at small driving potentials, it was demonstrated in Ref.~\cite{barnaveli2024asymmetric} that a lot of qualitative information can still be extracted even in the non-linear regime of interest here. At the same time, Ref.~\cite{barnaveli2024asymmetric} also showed that for low frequencies $\omega \tau_{s}\lesssim 1$, where $\tau_s=\sqrt{\tau_{RC}\tau_D}$, the circuit of Fig.~\ref{Fig:Equivalent_Circuit_Full} can be successfully approximated by a simplified circuit shown in Fig.~\ref{Fig:Equivalent_Circuit_Simplified}, which will be employed in this paper. Following the derivations in Ref.~\cite{barnaveli2024asymmetric}, and setting $C=C_1$, we first analytically calculate the charge $Q(t)$ accumulated in the capacitors of the circuit when the sawtooth driving voltage of Eq.~\eqref{eq:SawTooth} is applied, yielding
\begin{equation} \label{eq:ChargeDynamics}
	Q(t)=\frac{2 Q_0}{\pi} \sum_{n=1}^\infty \frac{(-1)^{n+1}}{n\sqrt{(n\omega RC)^2 +1}}\cos{(n\omega t+\varphi_n)},
\end{equation}
where $Q_0=\Psi_0 C$ is a reference charge and $\varphi_n=\arctan({1/(n\omega RC)})$ the $n$-the phase angle. In Fig.~\ref{Fig:ChargeDynamics} we plot two periods of $Q(t)/Q_0$ as a function of (dimensionless) time $t/T$ for the same driving as in Fig.\ref{Fig:SawTooth} (so for all harmonic modes rather than only two) for driving frequency $\omega \tau_{RC}=1$. The phase shift between voltage and charge is evident. The plot identifies the two (dimensionless) times $t_1$ and $t_2$ in between which $Q(t)>0$, and likewise the interval between $t_2$ and $t_3=t_1+1$ during which $Q(t)<0$. The plot also shows the maximum $q_1$, the minimum $q_2$, and the integrated (absolute) surface areas $S_3$ and $S_4$ under the curve of $Q(t)/Q_0$. We see for the present example that while the curve corresponding to the area $S_3$ has a higher amplitude than that of the area $S_4$, so $|q_1|>|q_2|$, the base of $S_4$ is actually wider, $\Delta t_1\equiv t_2-t_1<\Delta t_2 \equiv  t_3-t_2$. In the linear response regime this is such that $S_3=S_4$ when $S_1=S_2$ in Fig.~\ref{Fig:SawTooth}, which implies a vanishing period-averaged charge on the capacitor in this linearized case.

However, as we will see in more detail in Section~\ref{Sec:ParametricStudy} below, the electrolytic system of interest is in the non-linear screening regime with a nonzero period averaged (dimensionless) surface charge on the left electrode $\sigma' \sim \Psi_0^3$. This is a consequence of a nontrivial rescaling of the time-dependent electrode charge $\sigma(t)$, that causes the analogues of the extrema $q_1$ and $q_2$ of the charge curve to scale non-linearly with the voltage amplitude. In turn, this causes a nontrivial  relation between the amplitude difference $\Delta q \equiv |q_1|-|q_2|$ and the base width difference $\Delta t\equiv |\Delta t_1-\Delta t_2|$, leading to a non-zero time-averaged area $\Delta S=S_3-S_4\neq 0$ and consequently to a non-zero time-averaged surface charge $\sigma'$ with a sign that depends on the system parameters, as we will see in Section~\ref{Sec:ParametricStudy} below.   

\section{Parameter Dependence of AREF}
\label{Sec:ParametricStudy}

\begin{figure} [ht]
	\centering
	\includegraphics[width=0.99\linewidth]{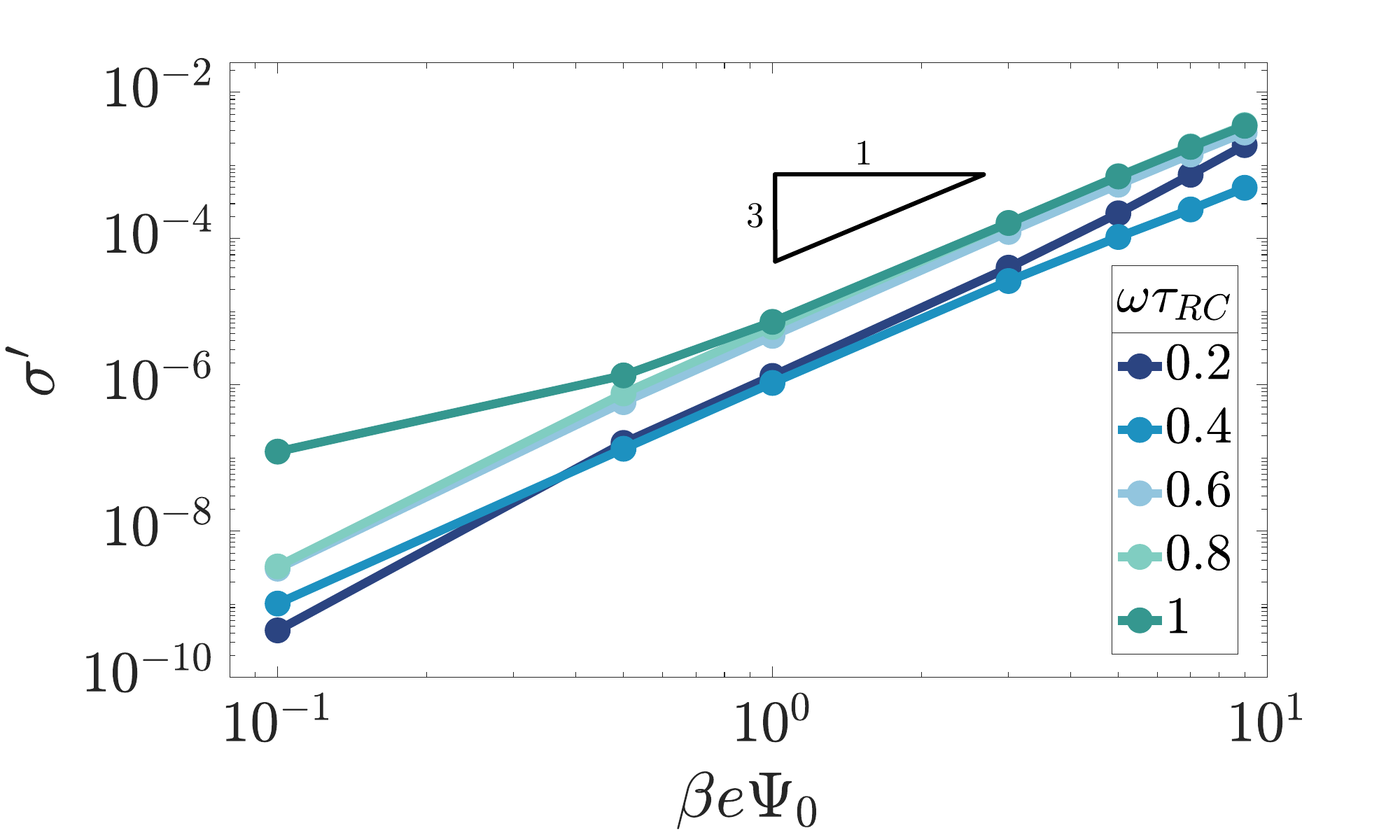}
	\caption{ Period-averaged dimensionless surface charge $\sigma'$ of Eq.~\eqref{eq:Sigma_Defined} plotted in the double-logarithmic representation against the driving voltage amplitude for varying driving frequencies $\omega$ at our standard parameter set (see text). The cubic scaling $\sigma'\sim \Psi_0^3$ demonstrates that AREF is a non-linear effect.} \label{Fig:Voltage_Dependence}
\end{figure}

\noindent In this section we study the dependence of the numerically obtained time-averaged surface charge $\sigma'$, defined in Eq.~\eqref{eq:Sigma_Defined},  on the main system parameters. We recall that
all numerical calculations are performed using the two-term truncation of Eq.~\eqref{eq:Sawtooth2Term}. The key results are presented in Fig.~\ref{Fig:Voltage_Dependence}, where we show that $\sigma'\propto\Psi_0^3$, and in Fig.~\ref{Fig:AREF_Parametric}, where we plot $\sigma'/(\beta e\Psi_0)^3$, in (a) and (c) as a function of the driving frequency for different driving amplitudes (a) and different phase angles $\Delta \phi$ between the two sinusoidal terms of the two-term sawtooth function in Eq.~\eqref{eq:Sawtooth2Term} (c) as we will see in more detail below, and in (b) as a function of system size at several driving frequencies. In all cases shown in Fig.~\ref{Fig:AREF_Parametric}, we see variations over an order of magnitude and even changes of the sign, which testify for the substantial tunability of AREF. However, we also see in Fig.~\ref{Fig:Voltage_Dependence} that the order of magnitude of $\sigma'$ is at most of the order of $10^{-3}$, such that the period-averaged surface charge $\langle\sigma\rangle$  is at least three orders of magnitude smaller than the typical static Gouy-Chapman surface charge density $\sigma_m$ at $\Psi_0=75~\text{mV}$ as defined below Eq.~\eqref{eq:Sigma_Defined} for our system parameters. This does not imply, however, that AREFs are a mere quantitative effect without qualitative consequences, since the force that is exerted by an AREF on a (colloidal) body also depends on its net charge (which should therefore be large enough for AREF to be physically relevant, we estimate typically three orders of magnitude larger than the unit charge for the present (typical) parameters). Therefore, we will investigate, discuss, and interpret the dependence of AREF on the system parameters in more detail below.

\subsection{Applied Voltage Amplitude}

Similarly to Ref.~\cite{barnaveli2024asymmetric}, the range that we consider for the driving voltage amplitude $\Psi_0$ is limited from above by the point ion approximation, which even for $c_s=1$mM can give rise to unrealistically high local concentrations within the point-ion limit due to strong ion crowding effects that take place in actual electrolytes at the electrodes \cite{kilic2007steric,fedorov2008towards,bazant2009towards}. This occurs beyond $\beta e \Psi_0\approx 8-9$, which is therefore the upper limit that we consider in Fig.~\ref{Fig:Voltage_Dependence}, where we plot, for various driving frequencies, the dependence of $\sigma'$ on $\Psi_0$ for our standard parameter set. The slope of the double-logarithmic curves is essentially identical to $3$ across the range of frequencies $\omega \tau_{RC}\in[0.2,1]$ that we consider here,  i.e. $\sigma'\propto\Psi_0^3$. This non-linear scaling confirms that AREF is a non-linear screening effect in the present case of a symmetric electrolyte driven by the sawtooth voltage, very similar to the earlier case of a sinusoidal voltage driving an asymmetric electrolyte as studied in Refs.~\cite{hashemi2018oscillating,hashemi2019asymmetric,barnaveli2024asymmetric}. This entices the further study of its dependence on frequency, 
the phase shift between the two harmonic modes of the driving voltage, and the system size in terms of the scaled form $ \sigma' /(\beta e\Psi_0)^3$ below.

\subsection{Frequency}
\begin{figure} [ht!]
\captionsetup[subfigure]{labelformat=empty}
\subfloat[\label{Fig:Frequency_Dependence}]{\subfigimg[width=\linewidth,pos=tl,vsep=4.3cm,hsep=1.25cm]{\large (a)}{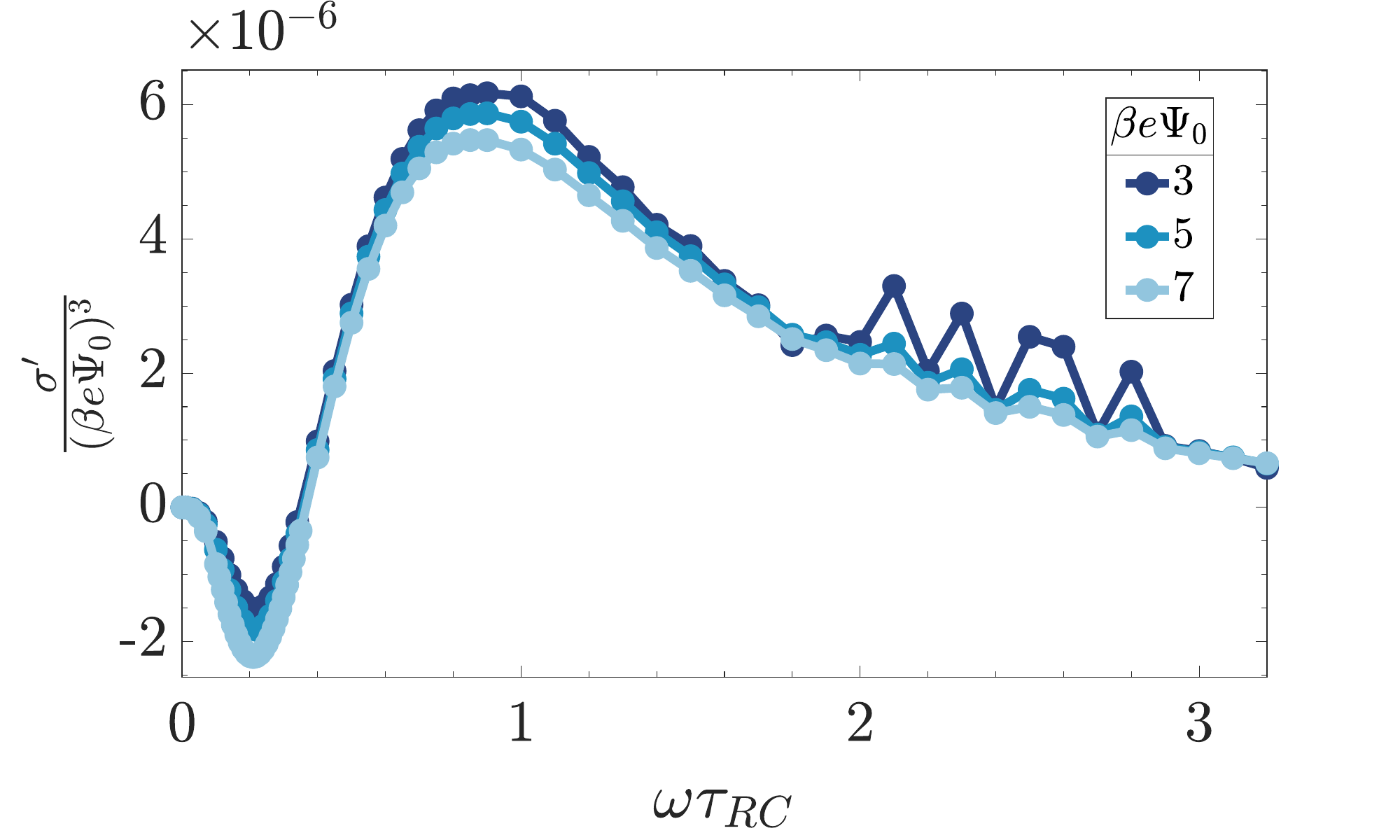}}\quad
\subfloat[\label{Fig:L_Dependence}]{\subfigimg[width=\linewidth,pos=tl,vsep=4.3cm,hsep=1.25cm]{\large (b)}{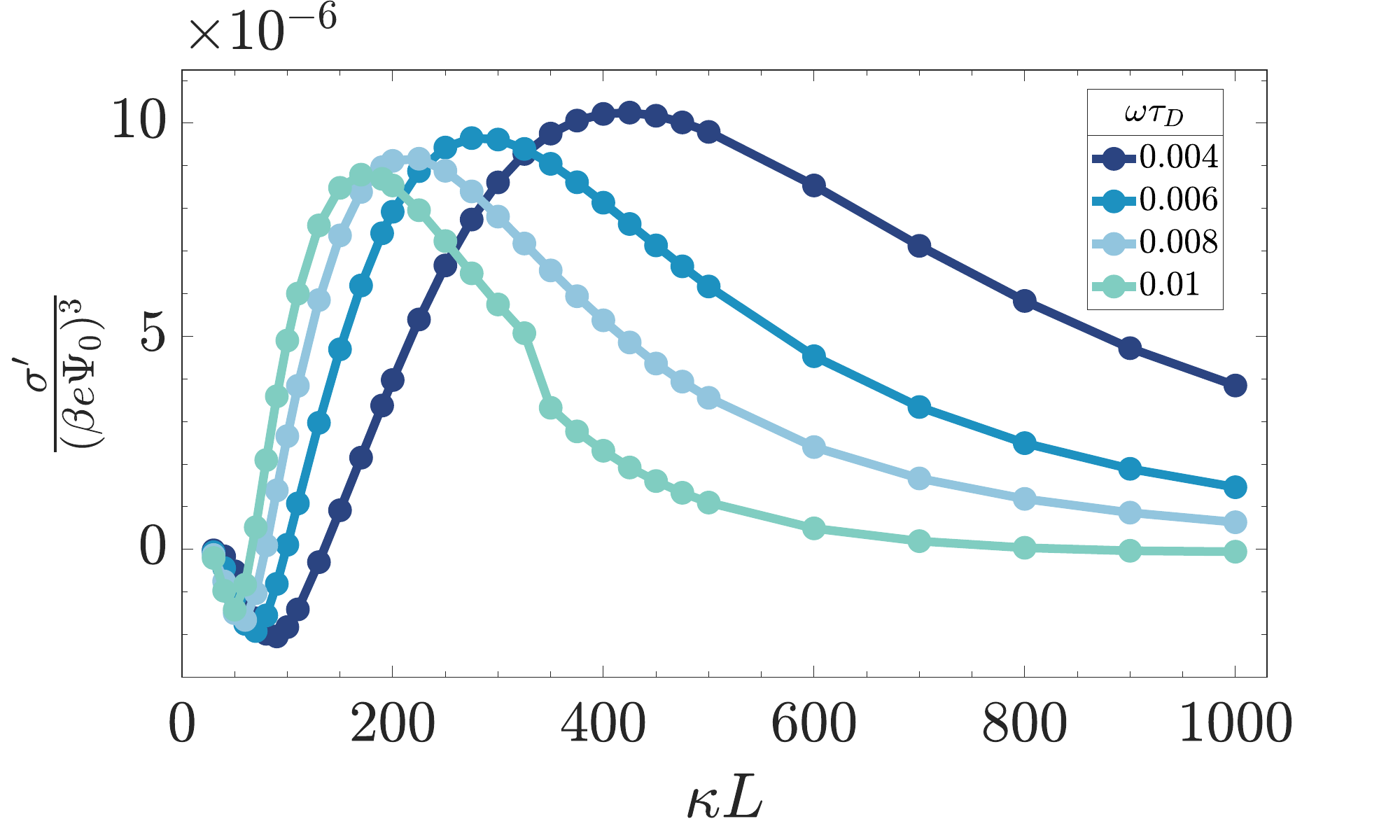}}\quad
\subfloat[\label{Fig:Frequency_Phase_Dependence}]{\subfigimg[width=\linewidth,pos=tl,vsep=4.3cm,hsep=1.55cm]{\large (c)}{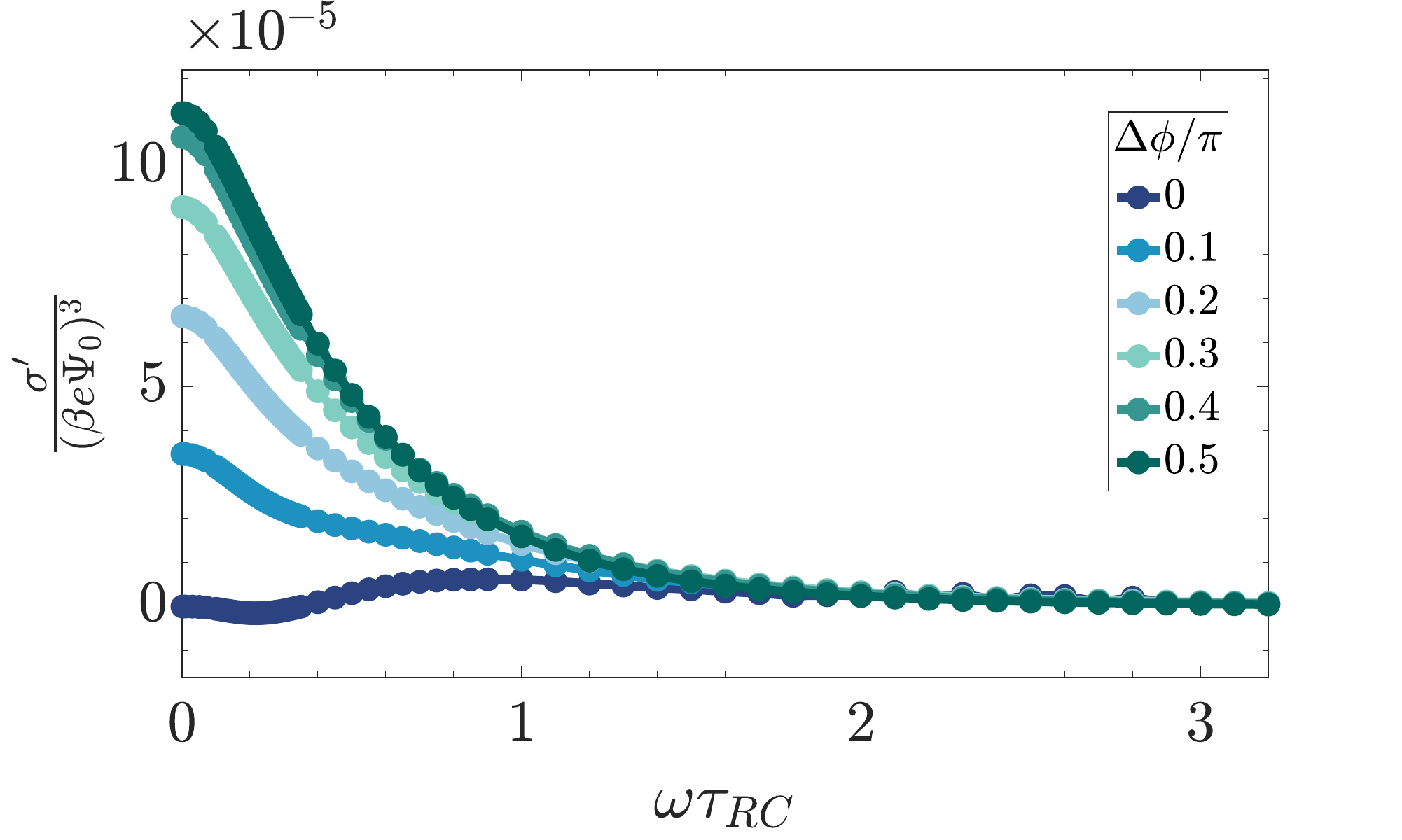}}
\caption{Numerically obtained period-averaged dimensionless surface charge $\sigma'/(\beta e \Psi_0)^3$ from late-time solutions of the PNP equations for the standard parameter set (see text) plotted against (a) the dimensionless frequency $\omega\tau_{RC}$ for several voltage amplitudes, (b) the dimensionless system size $\kappa L$ for several dimensionless driving frequencies $\omega\tau_D$ , and (c) the driving frequency $\omega\tau_{RC}$ for several phase shifts $\Delta \phi$ of Eq.~\eqref{eq:SawTooth2TermPhaseShift}.
In (a) we see a collapse of the curves for several voltage amplitudes $\Psi_0$.
\label{Fig:AREF_Parametric}}
\end{figure}

In Fig.~\ref{Fig:Frequency_Dependence} we plot $\sigma'/(\beta e\Psi_0)^3$ as a function of the dimensionless frequency $\omega \tau_{RC}$ for our standard parameter set at a number of voltage amplitudes $\Psi_0$. As expected, the curves essentially collapse for all $\Psi_0$ and decay to zero in the high and low frequency limits. We assign the irregularities in the graph for the lowest voltage in the high-frequency regime $\omega \tau_{RC}\sim 2-3$ as numerical artefacts without any significant physical meaning, stemming from the small numbers involved. Interestingly, however, in the frequency range $\omega \tau_{RC}\sim 0.1-2$ where the graphs are smooth, the average surface charge curves exhibit a change of sign while featuring both a positive maximum at $\omega \tau_{RC} \sim 1$ and a negative minimum at $\omega \tau_{RC} \sim 0.3$. The mechanism that generates such curves can be best understood in the context of an ``area competition'' between $S_3$ and $S_4$ under the $Q(t)$ curve for the equivalent circuit in Fig.~\ref{Fig:ChargeDynamics}, as we discussed above, but now with the time-dependent surface charge density $\sigma(t)$ obtained from the nonlinear PNP equations being the analogue of the capacitor charge $Q(t)$ in the linear circuit. 

\begin{figure} [ht]
	\centering
	\includegraphics[width=0.99\linewidth]{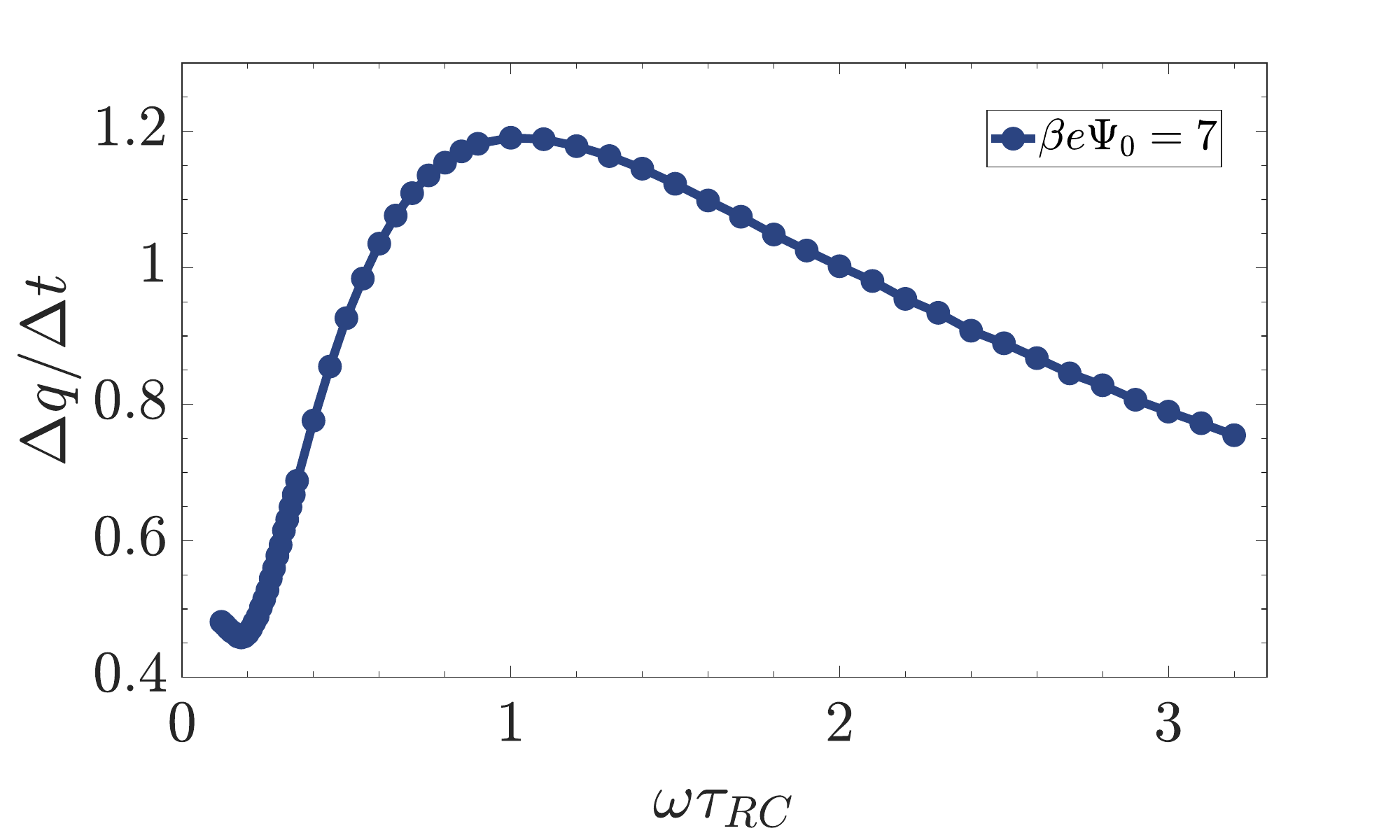}
	\caption{Ratio of the amplitude difference and time difference $\Delta q/\Delta t$ for the numerical solution $\sigma'$ as a function of dimensionless frequency $\omega \tau_{RC}$. Plotted for the standard parameter set (see text), however with $\beta e \Psi_0=7$ to minimize the numerical noise seen at higher frequencies in Fig.~\ref{Fig:Frequency_Dependence}. The shape of the $\Delta q/\Delta t(\omega)$ curve is remarkably similar to that of the $\sigma'(\omega)$ curve in Fig.~\ref{Fig:Frequency_Dependence}.} \label{Fig:DeltaPeak_DeltaTime}
\end{figure}

Depending on the parameter range, the $\sigma(t)$ analogue of either $\Delta q$ or $\Delta t$ dominates during a period of the (late time) voltage and charge oscillation, determining the the sign of the time-averaged charge. To check this statement, we calculate (the analogues of) $\Delta q$ and $\Delta t$ for the numerical results of $\sigma(t)$ (driven by the two-term sawtooth function) and plot their ratio $\Delta q/\Delta t$ as a function of the dimensionless frequency $\omega \tau_{RC}$ in Fig.~\ref{Fig:DeltaPeak_DeltaTime}. Interestingly, comparing this ratio to the $\sigma'(\omega)$ curve in Fig.~\ref{Fig:Frequency_Dependence}, we see a remarkable similarity in the shape of the curves, which suggests that a nontrivial competition between the amplitudes of the time-dependent surface charge and the duration of the time-interval of its positive and negative sign is indeed able to explain the nontrivial non-monotonic shape of the $\sigma'(\omega)$ curve of Fig.~\ref{Fig:Frequency_Dependence}.

\subsection{System Size}
Fig.~\ref{Fig:L_Dependence} shows the dependence of $\sigma'/(\beta e \Psi_0)^3$ on system size $L$ (in units of the Debye length) for various driving frequencies $\omega \tau_D$ for our standard parameter set. Rather than using the dimensionless combination $\omega \tau_{RC}$ of Eq.~\eqref{eq:tau_RC} to characterize the frequency of the driving voltage, here we use $\omega \tau_D$ defined in Eq.~\eqref{eq:tau_D} as this combination does not depend on $L$. The maximum $\sigma'$ for the relatively large system sizes of interest, say in the range of $\kappa L\in[10,10^3]$, occur at larger $\kappa L$ for lower frequencies $\omega\tau_D$, and one checks that they all correspond to the regime where $\omega \tau_{RC} \sim 1$. This agrees  with our findings of Fig.~\ref{Fig:Frequency_Dependence}. In fact, the dependence of $\sigma'$ on frequency in  Fig.~\ref{Fig:Frequency_Dependence} and on $L$ in Fig.~\ref{Fig:L_Dependence} are very similar, which in retrospect is not surprising since the key dimensionless parameter $\omega \tau_{RC}$ is linear in both $L$ and $\omega$.

\subsection{Phase Shift}
\begin{figure} [ht!]
\captionsetup[subfigure]{labelformat=empty}
\subfloat[\label{Fig:SawToothPhaseShift}]{\subfigimg[width=\linewidth,pos=tl,vsep=6.2cm,hsep=0.1cm]{\large (a)}{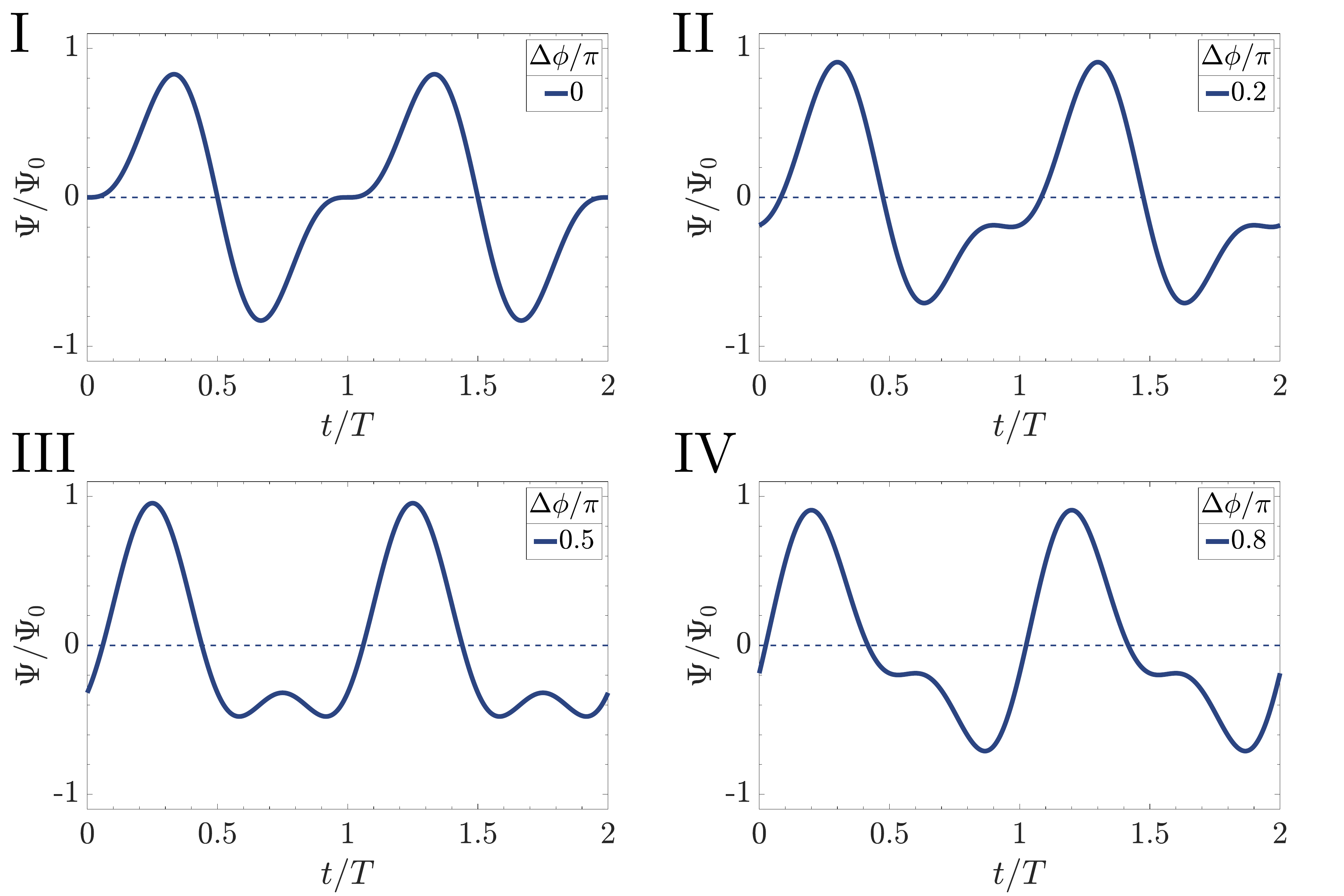}}\quad
\subfloat[\label{Fig:ChargePhaseShift}]{\subfigimg[width=\linewidth,pos=tl,vsep=6.2cm,hsep=0.1cm]{\large (b)}{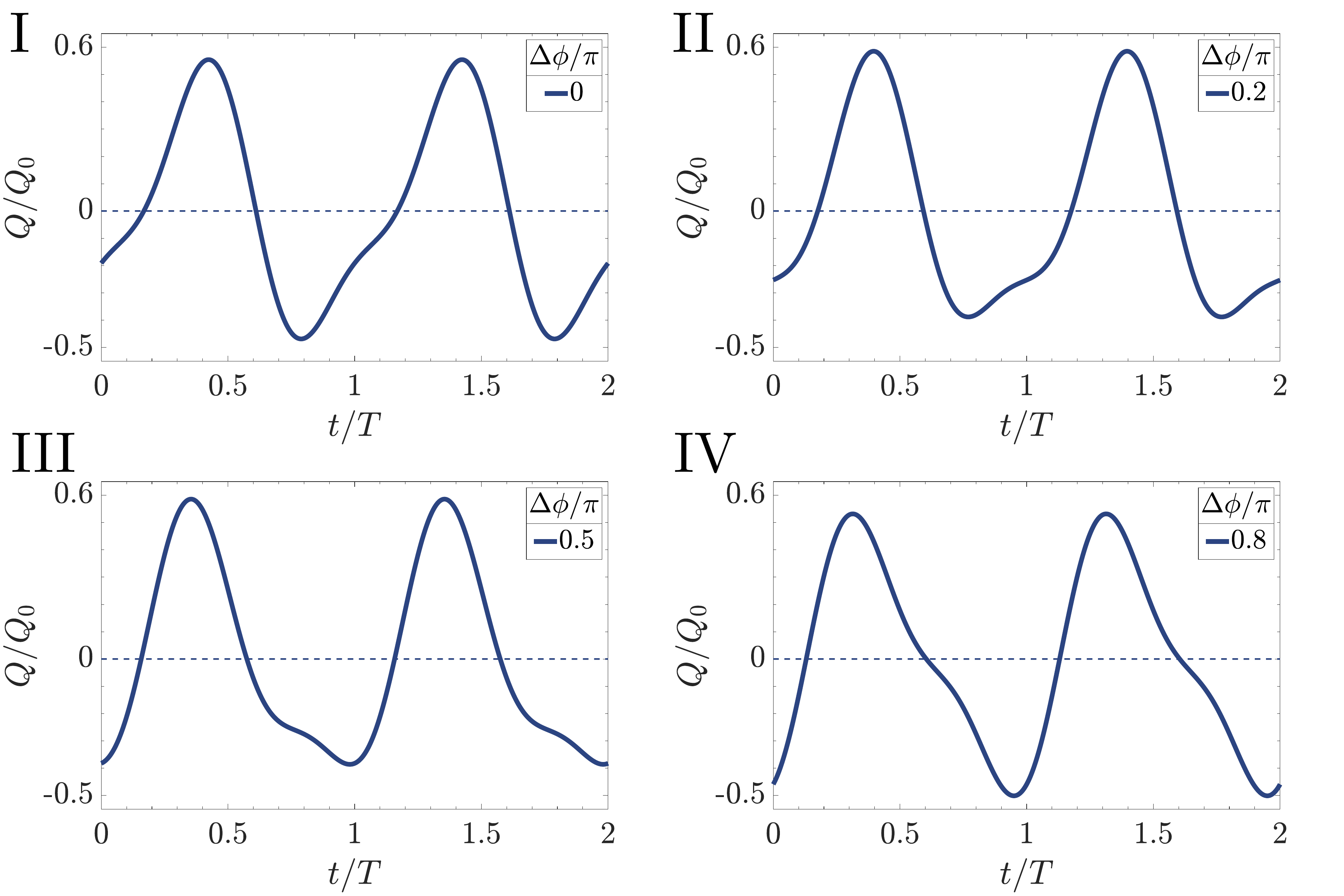}}
\caption{ (a) Two-term sawtooth voltage of Eq.~\eqref{eq:SawTooth2TermPhaseShift} for phase shifts $\Delta \phi /\pi$ equal to (I) $0$, (II) $0.2$, (III) $0.5$, and (IV) $0.8$, and (b) the resulting charges accumulating in the capacitors of the equivalent circuit of Fig.~\ref{Fig:Equivalent_Circuit_Simplified}. 
}
\end{figure}

As was mentioned in the introduction, the main advantage of using a sawtooth function to drive a symmetric electrolyte in the system of Fig.~\ref{Fig:TwoPlatesSetupAREF} compared to driving an asymmetric electrolyte with a sinusoidal voltage like in Ref.~\cite{barnaveli2024asymmetric}, is that one can manipulate AREF by simply altering the sawtooth potential without having to change the electrolyte properties (which would require the electrolyte to be changed in different experiments). As we are using the two-term sawtooth voltage of Eq.~\eqref{eq:AREF_Boundary_Conditions_Poisson_1}, it is thus interesting to see whether the AREF can be amplified or suppressed by shifting the relative phase $\Delta\phi$ between two sinusoidal terms away from zero. For this reason we consider the modified driving potential
\begin{equation} \label{eq:SawTooth2TermPhaseShift}
	{
		\Psi(t)=\frac{2\Psi_0 }{\pi}\bigg(\sin{(\omega t)}-\frac{1}{2}\sin{(2\omega t+\Delta \phi)}\bigg), 
	}
\end{equation}
which is identical to Eq.~\eqref{eq:Sawtooth2Term} for the case $\Delta\phi=0$. We note that a nonzero phase shift keeps the period-averaged driving potential equal to zero while it does affect the rate of voltage change substantially and the maximum/minimum voltage during a period somewhat. We plot this driving potential in Fig.~\ref{Fig:SawToothPhaseShift} at phase shifts $\Delta \phi /\pi =0, 0.2, 0.5$, and $0.8$ in the panels I through IV, respectively, together with the charge $Q(t)$ accumulated in the capacitor of the equivalent circuit of Fig.~\ref{Fig:Equivalent_Circuit_Simplified} in Fig.~\ref{Fig:ChargePhaseShift}. As we see in Fig.~\ref{Fig:SawToothPhaseShift}, any of the three nonzero phase shifts increases the maximum and decreases the minimum of the driving voltage, resulting in an increase of $\Delta q$ in the corresponding plots of $Q(t)$ in Fig.~\ref{Fig:ChargePhaseShift}. At the same time, while $\Delta t$ changes with $\Delta\phi$, it does not get affected by the non-linearity of AREF, thus it does not influence the surface charge dependence on the phase shift $\sigma'(\Delta\phi)$. On this basis, one could expect a strong effect of $\Delta\phi$ on the average surface charge $\sigma'$ in the non-linear electrolytic cell.

\begin{figure} [ht]
	\centering
	\includegraphics[width=0.99\linewidth]{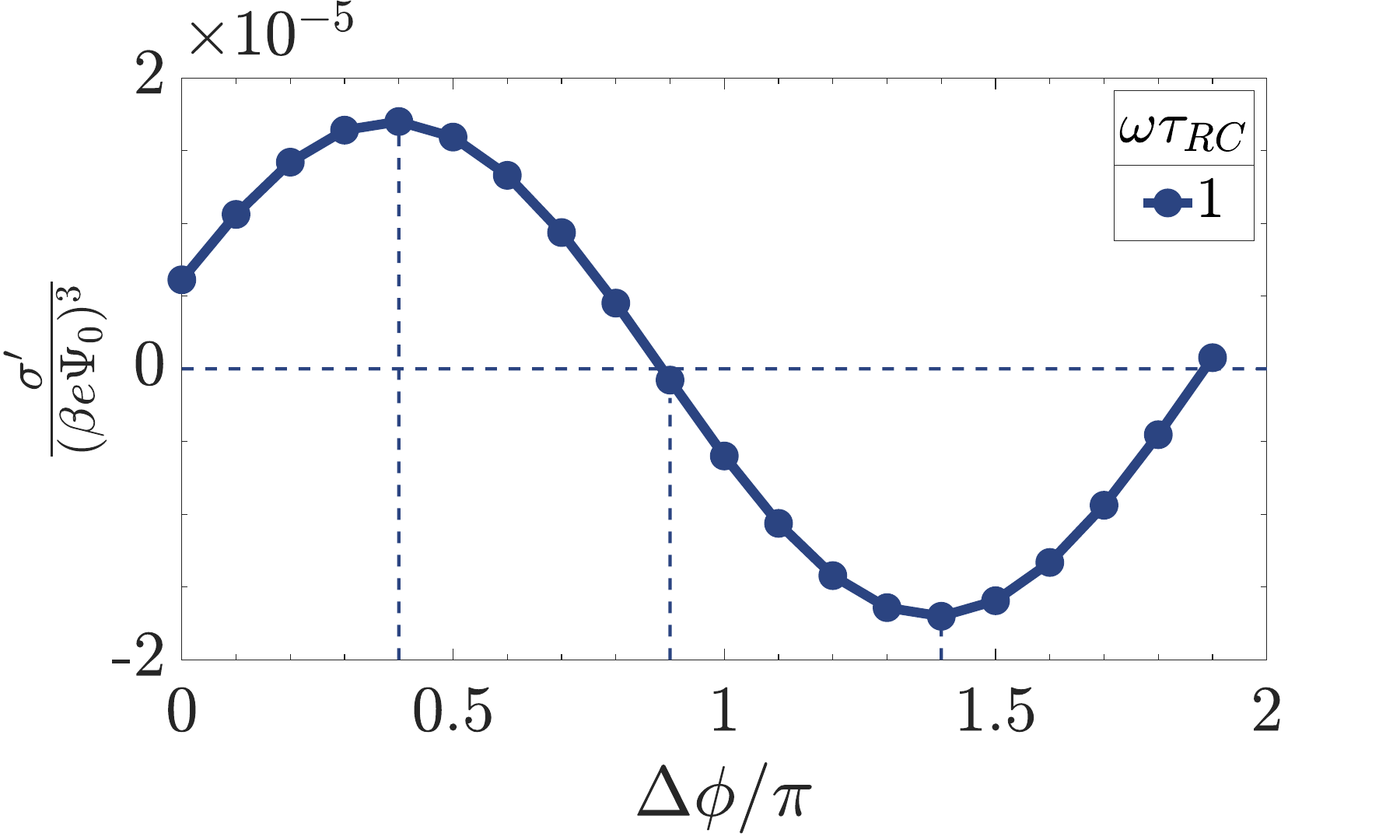}
	\caption{Dimensionless and scaled period-averaged surface charge $\sigma'/(\beta e \Psi_0)^3$ as obtained from numerical late-time solutions of the PNP equations for the standard parameter set (see text) as a function of the phase difference $\Delta \phi$ between the two sinusoidal terms of the two-term sawtooth potential of Eq.~\eqref{eq:Sawtooth2Term} at $\omega\tau_{RC}=1$. For these parameters the period-averaged surface charge has a maximum at $\Delta \phi=0.4\pi$ and a minimum (of opposite sign) at $\Delta \phi=1.4\pi$. }\label{Fig:Phase_Shift_Dependence}
\end{figure}

This strong effect of the phase shift is indeed confirmed by Fig.~\ref{Fig:Frequency_Phase_Dependence}, where we plot $ \sigma' /(\beta e\Psi_0)^3$ as a function of the dimensionless frequency $\omega \tau_{RC}$ for our standard parameter set at $\Delta \phi /\pi =0,0.2,0.5,$ and $0.8$. We see that as we shift the phase the AREF effect can actually increase by as much as an order of magnitude, reaching its highest values at $\Delta \phi=0.5\pi$. At the same time, we see that it only changes sign with frequency for the case $\Delta \phi=0$. Increase of $\Delta q$ with phase shift is well reflected in Fig.~\ref{Fig:Phase_Shift_Dependence}, where we plot $\sigma'$ as a function of the phase shift $\Delta \phi$ at a fixed frequency of $\omega \tau_{RC}=1$. 
As we see, the average surface charge has a maximum at $\Delta \phi=0.4\pi$ and a minimum at $\Delta \phi=1.4\pi$, where it also has the opposite sign.

\subsection{Sawtooth AREF vs. Symmetric AREF}

Here we briefly compare the spatial dependence and the magnitude of AREF in the present case of a symmetric electrolyte with equal ionic diffusion coefficients driven by a sawtooth voltage with the more conventional case of an asymmetric electrolyte (with different ionic diffusion coefficients) driven by a sinusoidal voltage. We focus on the period-averaged electric field profile shown in Fig.~\ref{Fig:AvgE} for the present standard parameter set and the equivalent plot shown in Fig.2(b) of Ref.~\cite{barnaveli2024asymmetric} for identical system parameters ($\beta e\psi_0=3$, $\omega\tau_{RC}=1$, $\kappa L=50$) at a ratio of ionic diffusion coefficients equal to 2 and 3.5. A striking difference, discussed briefly before, concerns the differences in mirror symmetry with respect to the midplane. Also, for the case of sawtooth driving we see two AREF peaks (a minimum and a maximum) of the same order of magnitude in the Debye-length vicinity of the electrodes, whereas in the case of the asymmetric electrolyte we only obtain a single peak (a minimum at one electrode and a maximum at the other in agreement with the mirror anti-symmetry). We also note that the scale of the AREF peaks is roughly an order of magnitude larger in the asymmetric case compared to the sawtooth case, however, the latter spreads almost twice as deep into the bulk of the electrolyte.

\section{Discussion and Conclusion}
\label{Sec:Conclusion}
In this work we investigate the time-averaged static electric field generated within the electrolytic cell depicted in Fig.~\ref{Fig:TwoPlatesSetupAREF} when exposed to a sawtooth-shaped AC potential, under the condition of equal diffusion coefficients for monovalent cations and anions, i.e., $D_+ = D_-$. We numerically solve the coupled non-linear Poisson-Nernst-Planck (PNP) equations for ionic  diffusion and migration in the cell to examine the dependence of the magnitude of the emerging asymmetric rectified electric field (AREF) on key system parameters. These parameters include the amplitude $\Psi_0$ of the applied AC sawtooth voltage, the driving frequency $\omega$, the phase shift $\Delta \phi$ between the lowest two harmonic modes of the driving potential, and the system size $L$, where we note that these system parameters can all be externally tuned without requiring a change of the electrolyte. 

The asymmetry in the rate of change of the driving sawtooth voltage induces, despite the equal diffusion coefficients of the cations and anions and despite a zero period-averaged applied voltage, a nonzero period-averaged electrode charge $\langle\sigma\rangle$ that is responsible for a nonzero period-averaged asymmetric rectified electric field (AREF) between the electrodes. While AREF fundamentally represents a non-linear screening phenomenon that we find to be proportional to $\Psi_0^3$, we could still obtain additional insights by conducting an analysis using the linear RC-circuit of Fig.~\ref{Fig:Equivalent_Circuit_Simplified} that was also used and derived in Ref.~\cite{barnaveli2024asymmetric}. The analytic expression for the time-dependent charge $Q(t)$ on the capacitor of this circuit, in particular the difference between (i) the maximum and the minimum of this charge (represented by $\Delta q$) and (ii) the duration of the time-interval of positive and negative charge (represented by $\Delta t$), provides a clue on the physics of the nonlinear phenomenon of AREF. These nonzero differences have opposite effects on the the period-averaged charge, which cancels identically even for nonzero $\Delta q$ and $\Delta t$ in the case of linear circuits. However, this cancellation is no longer exact in the nonlinear case of the PNP equations, where an intricate competition between $\Delta q$ (favoring a net positive charge for our parameter choices) and $\Delta t$ (favoring a net negative charge) depends sensitively on the system parameters. For driving frequencies $\omega$ that are of the same order as the inverse of the characteristic RC-time of electric double layers, i.e., when $\omega \tau_{RC}\sim 1$, this competition between $\Delta q$ and $\Delta t$ induces the most prominent period-averaged distribution of ionic charges, which, consequently, results in the largest non-zero AREF structure. The dependence on  the system size $L$ is largely reflected by the dependence on the RC-time -which also depends on $L$. A relatively strong AREF effect of an order of magnitude can be induced by a phase difference $\Delta\phi=\pi/2$ between the two modes of the driving voltage in the two-mode approximation. 

Finally, we noted that a recent investigation on floating colloids subjected to AC voltage within an electrolytic cell \cite{chen2023dielectrophoretic} proposed that apart from AREF also dielectrophoresis (DEP) might also play a role in counteracting the gravitational forces on the colloids, depending on the system parameters. However, the relative contribution of each of these mechanisms to the floating height of the colloids remains an open question. It may well be possible to separate the contributions of the two mechanisms by employing sawtooth potentials, which we have shown here offer substantial opportunities for tuning AREF without the need to change the electrolyte or the colloidal suspension. We hope that this work stimulates experimental work along these lines to manipulate a given electrolyte externally.

\section*{Conflicts of Interest}
There are no conflicts of interest to declare.

\begin{acknowledgments}
This work is part of the D-ITP consortium, a program of the Netherlands Organisation for Scientific Research (NWO) that is funded by the Dutch Ministry of Education, Culture and Science (OCW). We would like to thank Sanli Faez for his ideas and fruitful discussions on this topic and suggesting the sawtooth functional shape of the voltage.
\end{acknowledgments}

\section*{References:}
\nocite{*}
\bibliography{MS}

\end{document}